\documentclass[aps,prl,twocolumn,superscriptaddress,floatfix,letterpaper]{revtex4}
\usepackage{graphicx,epsfig,amsmath}
\newcommand{\subfig}[2]{Fig.~\ref{fig:#1}(#2)} %for instance \subfig{LargeDiamonds}{b}
\newcommand\degrees[1]{\ensuremath{#1^\circ}}

\newcommand{\Vone}{\mbox{$\text{V}_{\text{1}}$}}
\newcommand{\Vtwo}{\mbox{$\text{V}_{\text{2}}$}}

\newcommand{\NL}{\mbox{$\text{N}_{\text{L}}$}}
\newcommand{\NC}{\mbox{$\text{N}_{\text{C}}$}}
\newcommand{\NR}{\mbox{$\text{N}_{\text{R}}$}}
\newcommand{\Tts}{\mbox{$\text{T}_{\text{2}}^{\text{*}}$}}
\newcommand{\Tzero}{\mbox{$\text{T}_{\text{0}}$}}
\newcommand{\Tplus}{\mbox{$\text{T}_{\text{+}}$}}
\newcommand{\Tone}{\mbox{$\text{T}_{\text{1}}$}}

\newcommand{\Qtt}{\mbox{$\text{Q}_{\text{3/2}}$}}
\newcommand{\Qit}{\mbox{$\text{Q}_{\text{1/2}}$}}
\newcommand{\Dit}{\mbox{$\Delta_{1/2}$}}
\newcommand{\Dpit}{\mbox{$\Delta '_{1/2}$}}

\begin{document}

% The title
%\title{Coherent spin manipulation in a triple quantum dot}
\title{Coherent control of three-spin states in a triple quantum dot}

% The authors

\author{L.~Gaudreau}
	\altaffiliation{These authors had equal contributions to this work.}
	\affiliation{Institute for Microstructural Sciences, National Research Council Canada, Ottawa, ON Canada K1A 0R6}
	\affiliation{D\'epartement de physique, Universit\'e de Sherbrooke, Sherbrooke, QC Canada J1K 2R1}
\author{G.~Granger}
	\altaffiliation{These authors had equal contributions to this work.}
	\affiliation{Institute for Microstructural Sciences, National Research Council Canada, Ottawa, ON Canada K1A 0R6}
\author{A.~Kam}
	\affiliation{Institute for Microstructural Sciences, National Research Council Canada, Ottawa, ON Canada K1A 0R6}
\author{G.~C.~Aers}
	\affiliation{Institute for Microstructural Sciences, National Research Council Canada, Ottawa, ON Canada K1A 0R6}
\author{S.~A.~Studenikin}
	\affiliation{Institute for Microstructural Sciences, National Research Council Canada, Ottawa, ON Canada K1A 0R6}
\author{P.~Zawadzki}
	\affiliation{Institute for Microstructural Sciences, National Research Council Canada, Ottawa, ON Canada K1A 0R6}
\author{M.~Pioro-Ladri\`ere}
	\affiliation{D\'epartement de physique, Universit\'e de Sherbrooke, Sherbrooke, QC Canada J1K 2R1}
\author{Z.~R.~Wasilewski}
	\affiliation{Institute for Microstructural Sciences, National Research Council Canada, Ottawa, ON Canada K1A 0R6}
\author{A.~S.~Sachrajda}
  \email{Andrew.Sachrajda@nrc.ca}
	\affiliation{Institute for Microstructural Sciences, National Research Council Canada, Ottawa, ON Canada K1A 0R6}

% date
%\date{\today}

% The abstract
%\begin{abstract}
%We experimentally demonstrate Landau-Zener-St\"uckelberg (LZS) oscillations in a triple quantum dot environment. Using a pulsing technique in the spin qubit regime, we create a superposition of three-electron states, allow for phase accumulation, and readout. We demonstrate coherent LZS oscillations with three spins across the triple quantum dot structure and investigate, both experimentally and numerically, their dependence on pulse duration, detuning, magnetic field, and pulse rise time. The interplay between two qubits from this system is also observed and modelled.
%\end{abstract}

% The PACS:

% make the title
\maketitle

% begin the article:	

%\section{Introduction}

% Put here Andy's Word version.

\textbf{Spin qubits involving individual spins in single quantum dots (QDs) or coupled spins in double quantum dots (DQDs) have emerged as potential building blocks for quantum information processing applications\cite{Petta2005, Koppens2006, Hanson2007, PioroLadriere2008}. It has been suggested that triple quantum dots (TQDs) may provide additional tools and functionalities. These include the encoding of information to either obtain protection from decoherence or to permit all-electrical operation \cite{DiVincenzo2000}, efficient spin busing across a quantum circuit \cite{Greentree2004}, and to enable quantum error correction utilizing the three-spin Greenberger-Horn-Zeilinger quantum state.  Towards these goals we demonstrate for the first time coherent manipulation between two interacting three-spin states. We employ the Landau-Zener-St\"uckelberg \cite{Shevchenko2010, Zener1932} (LZS) approach for creating and manipulating coherent superpositions of quantum states \cite{Petta2010}. We confirm that we are able to maintain coherence when decreasing the exchange coupling of one spin with another while simultaneously increasing its coupling with the third. Such control of pairwise exchange is a requirement of most spin qubit architectures \cite{Loss1998} but has not been previously demonstrated.}

\begin{figure}[hbt]
\setlength{\unitlength}{1cm}
\begin{center}
\begin{picture}(8,6.7)(0,0)
\includegraphics[width=8cm, keepaspectratio=true]{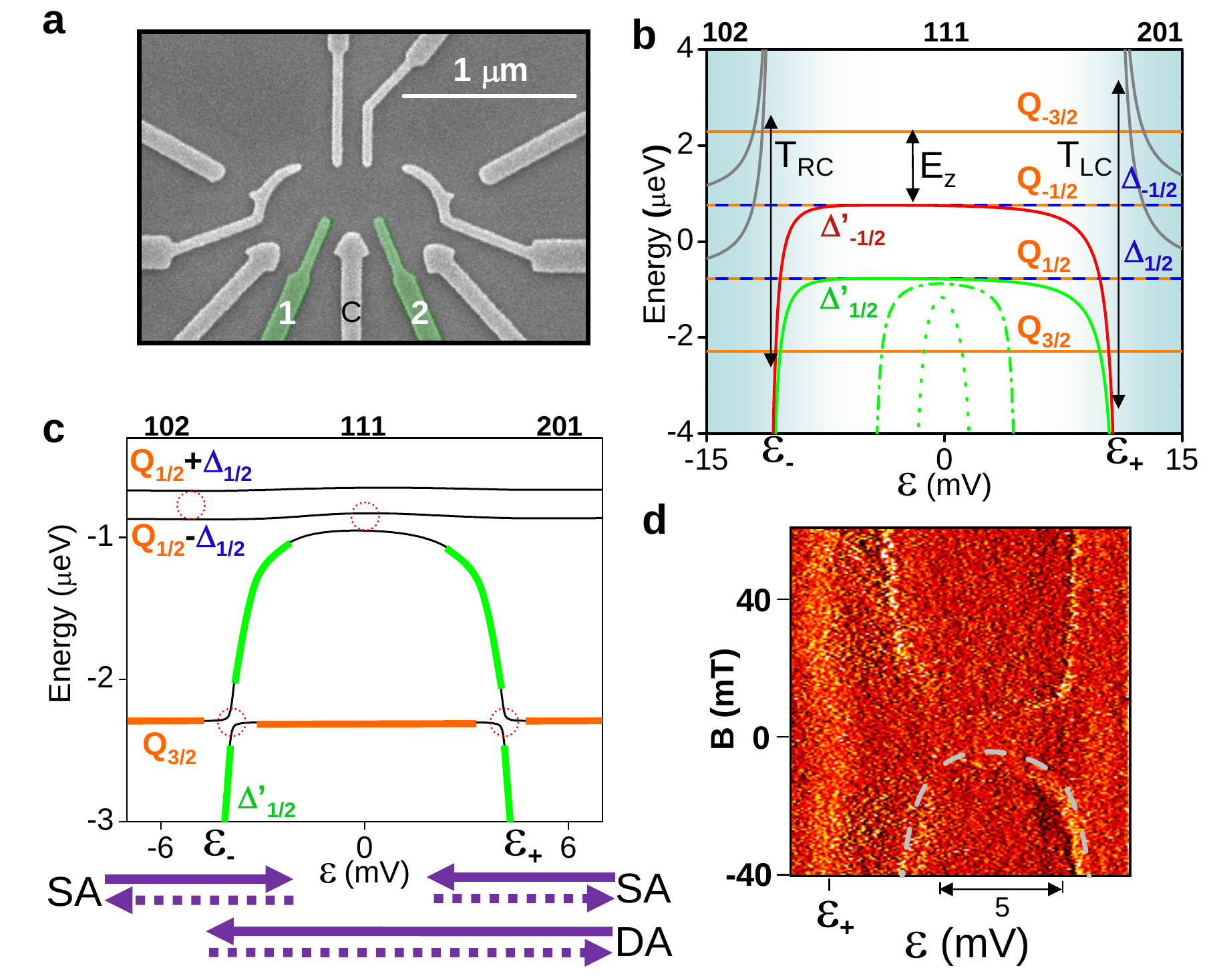}
\end{picture}
\end{center}
\caption{\footnotesize{\textbf{Device, three-spin states spectrum, and spin arch.} (a) Electron micrograph of a device identical to the one measured. Gates 1 and 2 are connected to high frequency lines for the application of fast voltage pulses ($\delta V_1$,$\delta V_2$) in addition to DC voltages (\Vone,\Vtwo). Gate C tunes the (1,1,1) region size by shifting the centre dot addition line. (b) Calculated energies vs.~detuning $\epsilon$ for the three-spin states for a 22-mV-wide (1,1,1) region (i.e.~$\left|\epsilon_+-\epsilon_-\right|$=22~mV), neglecting the hyperfine interaction. The Zeeman splitting, E$_z$, originates from an applied 60~mT field. The detuning line is describing a \degrees{-45} angle with respect to the \Vone~axis in the \Vone-\Vtwo~plane. The states shown in grey are split by the tunnel couplings $T_{RC}$ and $T_{LC}$ (not drawn to scale) from the $\Delta'$ states. The \Dpit~state is also drawn for a midsized (1,1,1) region (green dash-dotted line) and for a narrow (1,1,1) region (green dotted curve). (c) Calculated energy diagram including the effect of hyperfine interaction resulting from the proximity of the four lowest energy three-spin states with $S_z>0$ (states with S$_z<$0 are excluded for simplicity). Dotted red circles indicate pairs of states coupled by the hyperfine interaction. The dotted red circle at $\epsilon$=0 represents the hyperfine interaction between \Dpit~and \Qit~(the meaning of the remaining dotted red circles is clear). The single \Dpit-\Qtt~anticrossing (SA) and double \Dpit-\Qtt~anticrossing (DA) pulses are drawn. (d) Numerical derivative of the left QPC conductance with respect to \Vtwo~in the presence of a pulse across the charge transfer line between (2,0,1) and (1,1,1) for a 9-mV-wide (1,1,1) region . The extent of the (1,1,1) region along the detuning line (approximately joining the centers of the two charge transfer lines) is measured by a projection onto the gate voltage axis that is on the same side as the QPC detector used in the measurement. It is the resulting gate voltage range that is set equal to $|\epsilon_+ - \epsilon_-|$, and this is used for comparison between regimes of (1,1,1) regions with different widths. Black is low, orange is medium, and yellow is high. The pulse shape is in the Supplementary Information. The detuning line makes a \degrees{-51.3} angle with respect to the \Vone~axis in the \Vone-\Vtwo~plane, permitting both sides of the spin arch to be observed. The dashed line is the theoretical fit (with detuning-dependent interdot couplings included).}}
\label{fig:1}
\end{figure}

\begin{figure}[hbt]
\setlength{\unitlength}{1cm}
\begin{center}
\begin{picture}(8,5.8)(0,0)
\includegraphics[width=8.0cm, keepaspectratio=true]{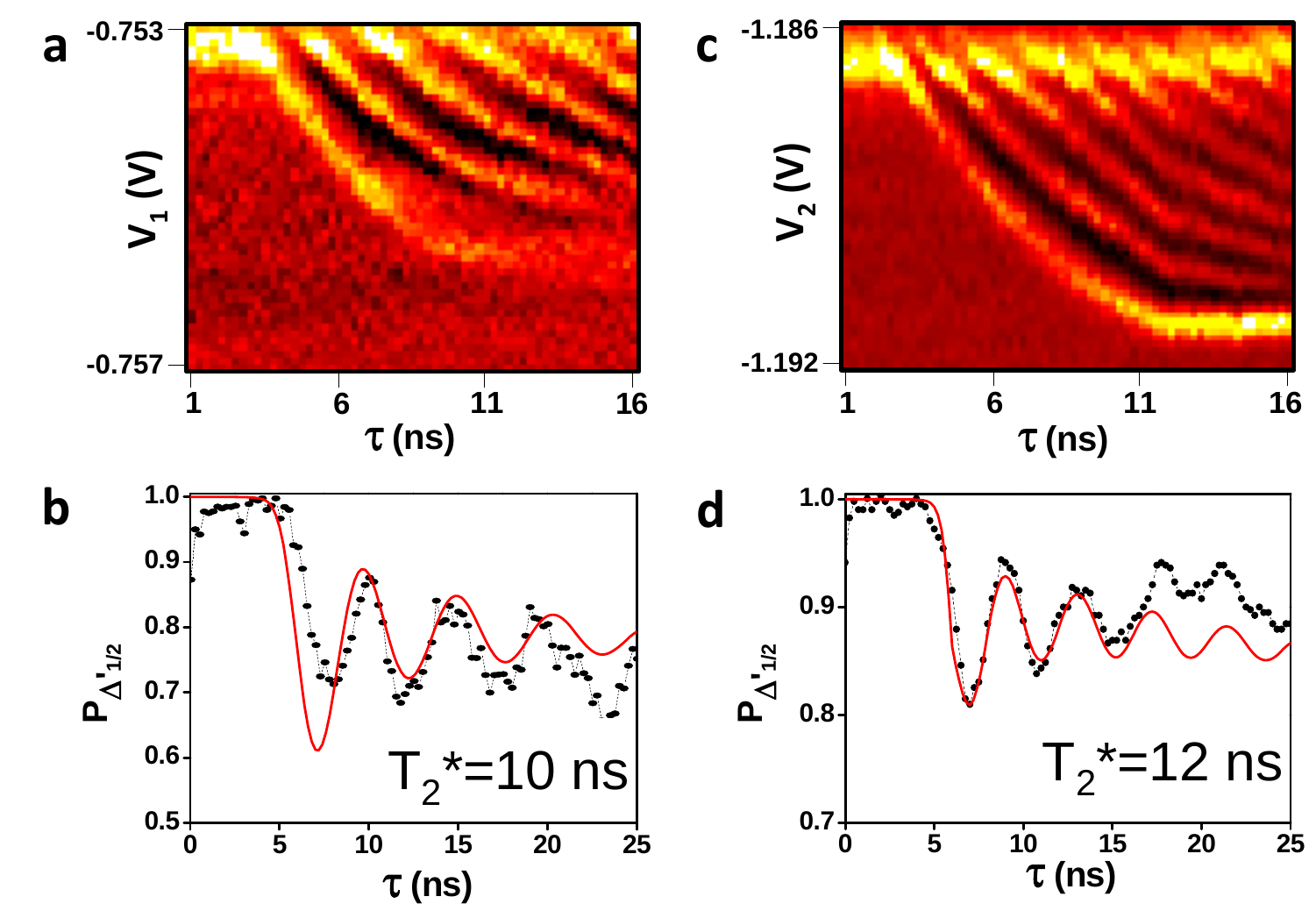}
\end{picture}
\end{center}
\caption{\footnotesize{\textbf{LZS oscillations from the two \Dpit-\Qtt~qubits for a wide (1,1,1) region.} The data in figure (a) and (b) are taken with the right QPC and (c) and (d) with the left QPC. (a) [c] Numerical derivative of the right [left] QPC conductance with respect to detuning along \Vone~[\Vtwo] illustrating LZS oscillations vs.~pulse duration $\tau$ across the (1,0,2) [(2,0,1)] to (1,1,1) charge transfer line at B=60~mT. Black is low, red is medium, and yellow is high. $|\epsilon_+ - \epsilon_-|$=27~mV along \Vone~[$|\epsilon_+ - \epsilon_-|$=41.5~mV along \Vtwo]. In (a), both \Vtwo~and \Vone~are swept in order to detune parallel to the pulse direction in the \Vone-\Vtwo~plane. (b and d) Probability of ending in the \Dpit~state as a function of $\tau$ with fits for \Tts. (b) [d] The pulse goes from (1,0,2) [(2,0,1)] to (1,1,1) and $|\epsilon_+ - \epsilon_-|$$\sim$50~mV along \Vone~[$|\epsilon_+ -\epsilon_-|$=27~mV along \Vtwo]. The experimental data are shown as points, while the theoretical fits are shown as red lines. The values of \Tts~extracted from the single parameter fit to the LZS model are indicated.}}
\label{fig:2}
\end{figure}

\begin{figure}[thb]
\setlength{\unitlength}{1cm}
\begin{center}
\begin{picture}(8,6.3)(0,0)
\includegraphics[width=8cm, keepaspectratio=true]{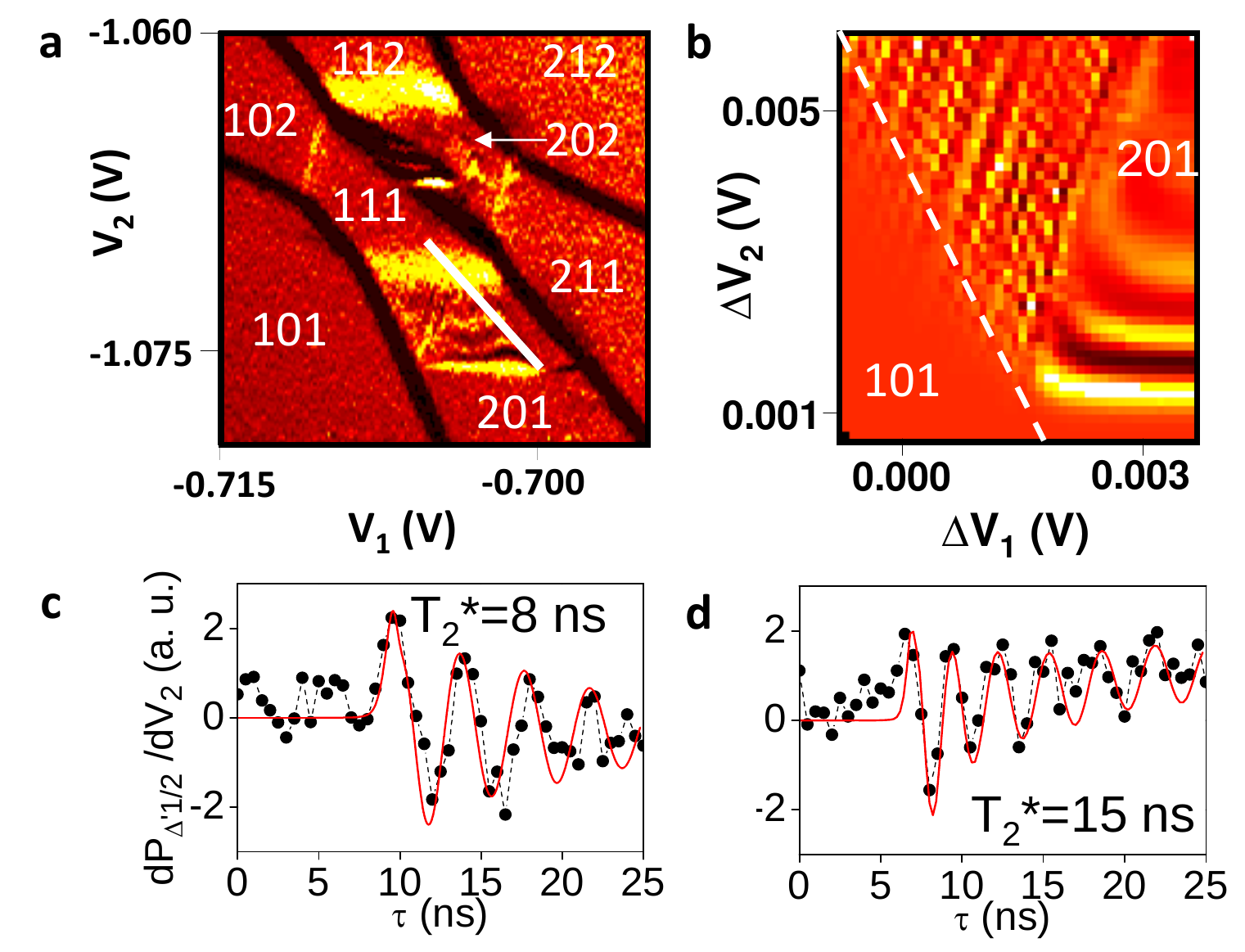}
\end{picture}
\end{center}
\caption{\footnotesize{\textbf{Coherent three-spin state manipulation with a narrow (1,1,1) region} (a) Stability diagram in the presence of a pulse [drawn as a white line for a given (\Vone,\Vtwo)], showing coherent LZS oscillations in the (2,0,1) region with features parallel to both charge transfer lines. The color map (black is low, red is medium, and yellow is high) corresponds to the numerical derivative of the left QPC conductance with respect to \Vtwo~in the presence of a pulse across the charge transfer line between (2,0,1) and (1,1,1). The (1,1,1) region is tuned to a width of $\sim$5~mV with gate C. B=25~mT. The stability diagram also shows LZS oscillations involving (2,0,2) and (1,1,2). (b) Calculated dP$_{\Delta'_{1/2}}$/d\Vtwo~map zooming mainly into the (2,0,1) region of the stability diagram from (a). The dashed line shows where the addition line is expected, although it is not part of the calculation. B=40~mT.  (c and d) Traces of dP$_{\Delta'_{1/2}}$/d\Vtwo~vs.~$\tau$. The data points in (c) [d] are extracted from \subfig{4}{b} (40~mT) at \Vtwo=-1.0751~V (white line) [-1.074 V (blue line)]. The fits (red lines) use B=60~mT. The values of \Tts~extracted from the fits are indicated.}}
\label{fig:3}
\end{figure}

\begin{figure*}[thb]
\setlength{\unitlength}{1cm}
\begin{center}
\begin{picture}(16,12)(0,0)
\includegraphics[width=16cm, keepaspectratio=true]{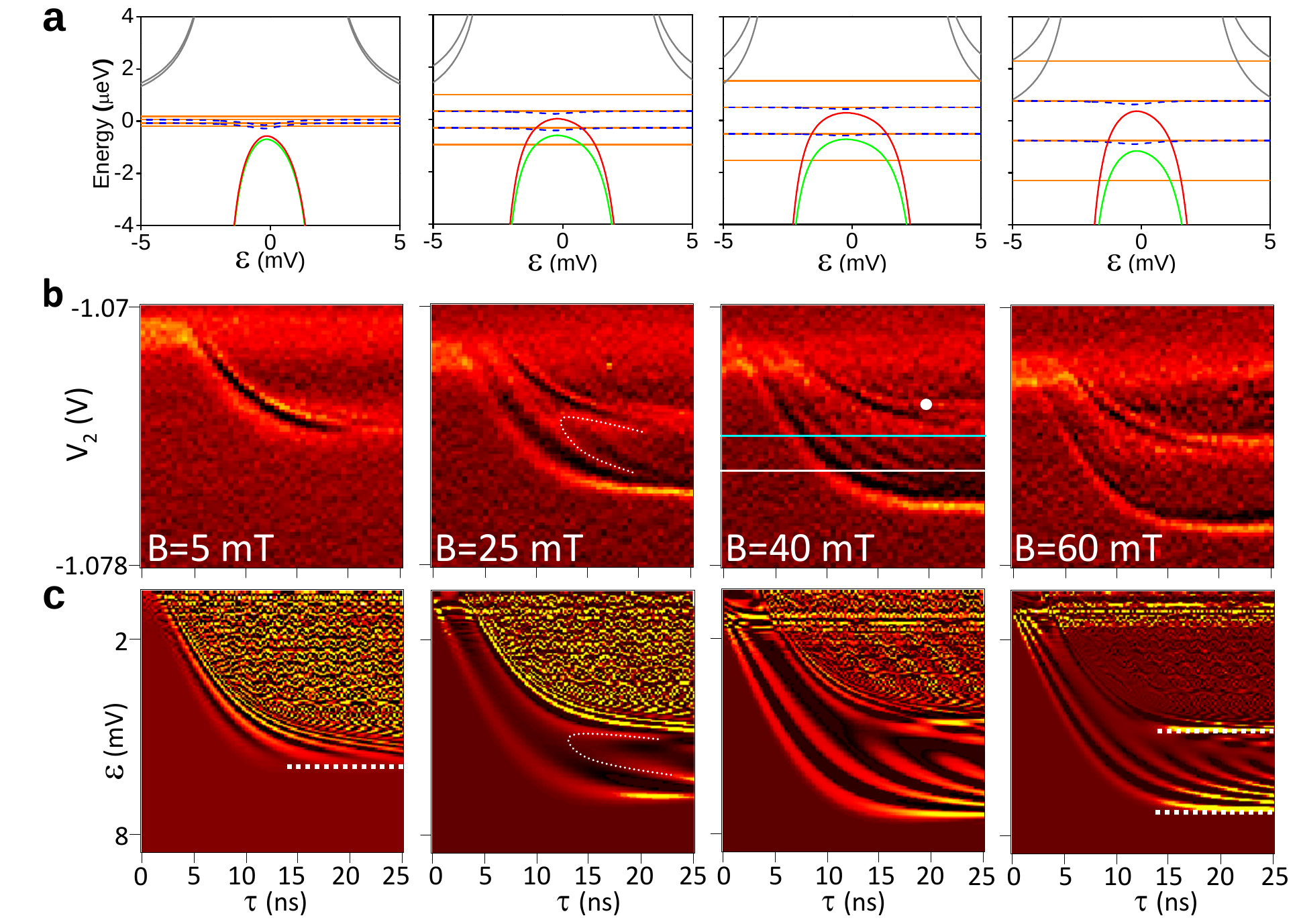}
\end{picture}
\end{center}
\caption{\footnotesize{\textbf{Magnetic field dependence of coherent three-spin state manipulation with a narrow (1,1,1) region.} (a) Energy spectra for the three-spin states for different magnetic fields. The color code for the states is the same as in \subfig{1}{b}. From left to right we have: B=5~mT and $\left|\epsilon_+-\epsilon_-\right|$=3.9~mV; B=25~mT and $\left|\epsilon_+-\epsilon_-\right|$=5.1~mV; B=40~mT and $\left|\epsilon_+-\epsilon_-\right|$=5.6~mV; and B=60~mT and $\left|\epsilon_+-\epsilon_-\right|$=4.6~mV. (b) Coherent oscillations shown in the $\tau$-\Vtwo~plane as the numerical derivative of the left QPC conductance with respect to \Vtwo~(black is low, red is medium, and yellow is high) in the presence of a pulse across the charge transfer line between (2,0,1) and (1,1,1). The (1,1,1) region is tuned to a width of $\sim$5~mV with gate C. \Vone~is swept proportionally to \Vtwo~in order to detune parallel to the pulse direction. The magnetic field and (1,1,1) region sizes from left to right are as in (a). The white dot in the B=40~mT map indicates a coherent oscillation resulting from a DA pulse reaching past the far \Dpit-\Qtt~anticrossing. (c) Calculated dP$_{\Delta'_{1/2}}$/d\Vtwo~maps (black is low, red is medium, and yellow is high) in the $\tau$-$\epsilon$~plane for the same experimental settings as for the panels in (b). The magnetic field and (1,1,1) region sizes from left to right are as in (a). No dephasing is included to keep the fringes clearer. The very rapid oscillations in the upper right corner of the figures are an artefact due to the large exchange energy past the far \Dpit-\Qtt~anticrossing. At B=5~mT, the anticrossings have merged, so there is only one boundary in the diagram (white dashed line). At B=60~mT, two \Dpit-\Qtt~anticrossings are recovered (see the two white dashed lines). At B=25~mT, for both theory and experiment, dotted white curves are drawn as a guide to the eye for the peak of an oscillation in between the two \Dpit-\Qtt~anticrossings. We note that a small dynamical nuclear polarisation (DNP) effect \cite{Brataas2011} is present which depends on the size of the (1,1,1) region, details of pulse shape, and pulse orientation. In (b) and (c) (25 and 40 mT), it is found that a DNP $\sim$20~mT is required to properly describe the period of oscillations. This is why the stability diagram in \subfig{3}{b} is calculated at 40~mT rather than 25~mT, and why the fits in \subfig{3}{c,d} are calculated at 60~mT instead of 40~mT.}}
\label{fig:4}
\end{figure*}

Following the spin qubit proposal by Loss and DiVincenzo \cite{Loss1998} and the electrostatic isolation of single spins in QDs \cite{Ciorga2000} and DQDs \cite{Elzermann2003}, coherent manipulation was demonstrated in two-level systems based on single-spin up and down states \cite{Koppens2006} as well as two-spin singlet and triplet states \cite{Petta2005}. Here we demonstrate coherent manipulation of a two-level system based on three-spin states. We employ the TQD device layout shown in \subfig{1}{a} consisting of multiple metallic gates on a GaAs/AlGaAs heterostructure. The gates are used to electrostatically define three QDs in series within a two-dimensional electron gas 110~nm below the surface. The QDs are surrounded by two quantum point contact charge detectors (QPCs) \cite{Field1993}. The QPC conductance identifies the number of electrons in each QD and its derivative with respect to a relevant gate voltage maps out the device configuration stability diagram. We tune the device to the qubit operating electronic configuration, (\NL,\NC,\NR)=(1,1,1), between two spin-to-charge conversion regimes (1,0,2) and (2,0,1), where L, C, and R refer to the left, centre, and right QDs respectively. The detuning, $\epsilon$, controls the energy difference between configurations (1,0,2), (1,1,1), and (2,0,1). The exchange coupling, $J$, depends upon $\epsilon$ and the tunnel couplings. 

In this paper we concentrate on two scenarios. In the first one, at each point in the stability diagram the exchange coupling to the centre spin from one or both of the edge spins is minimal (i.e.~one edge spin resembles a passive spectator). This configuration is used as a control to confirm that our device maps onto two-spin results in this limit \cite{Petta2010}. In the second scenario a true three-interacting-spin regime is achieved. (Results from a third intermediate regime are shown in the Supplementary Information.)

The energy level spectrum of a TQD \cite{Laird2010} consists of quadruplets Q with total spin $S$=3/2 separated by the Zeeman energy in a magnetic field and doublets $\Delta'$ and $\Delta$ with $S$=1/2. The two states of our qubit consist of one of the quadruplets, \Qtt, and one of the doublets, \Dpit, where
\begin{eqnarray}
&\lvert Q_{3/2}\rangle = \lvert \uparrow \uparrow \uparrow \rangle \nonumber \\
&\lvert \Delta'_{1/2}\rangle =
\frac{(-J_{LC}+J_{RC}+\Omega)\lvert\uparrow\uparrow
\downarrow\rangle-(J_{RC}+\Omega)\lvert\uparrow\downarrow\uparrow\rangle
+J_{LC}\lvert\downarrow\uparrow\uparrow\rangle
}{\sqrt{4{\Omega}^2+2
\Omega(2J_{RC}-J_{LC})}} \nonumber
\end{eqnarray}
with $\Omega=\sqrt{J_{LC}^2+J_{RC}^2-J_{LC}J_{RC}}$, and where $J_{LC}$  ($J_{RC}$) is the exchange coupling between the  left (right) and centre spins. (Other three-spin states are described in more detail in the Supplementary Information.)

Figure~\ref{fig:1}(b) illustrates the three-spin energy spectrum as a function of detuning [zero detuning is defined as the centre of the (1,1,1) regime as shown]. Experimentally we can tune the (1,1,1) region size by using gate C primarily \cite{Granger2010}. The eigenvalues of the four lowest states relevant for our experiments are:
\begin{eqnarray}
E_{\Qit} &=& -E_Z/2 \nonumber \\  
E_{\Dit} &=& -(J_{LC}+J_{RC}-\Omega+E_Z)/2 \nonumber \\ 
E_{\Qtt} &=& -3E_Z/2 \nonumber \\ 
%E_{\Delta} &=& -(J_{LC}+J_{RC}-\Omega)/2-E_ZS_{\Delta}^z\\
E_{\Dpit} &=& -(J_{LC}+J_{RC}+\Omega+E_Z)/2 \nonumber
\end{eqnarray}

The hyperfine interaction \cite{Taylor2007} couples the state  \Dpit~to the state \Qtt~(\Qit) at their anticrossing (asymptotic approach), see \subfig{1}{c}. (\Qit~and  \Dit~are also hyperfine coupled.) Figure~\ref{fig:1}(c) also illustrates the two types of experiment we describe in this paper. With the single anticrossing (SA) pulse based upon the methodology in Ref.~\cite{Petta2010}, the system starts in the  \Dpit~state in the (2,0,1) region [or (1,0,2)] and then a pulse is applied to reach the (1,1,1) regime. The pulse rise time (see supplementary information) ensures that Landau-Zener (LZ) tunneling creates a coherent superposition of \Qtt~and \Dpit~on passage through the anticrossing. After a state evolution time, $\tau$, the pulse steps down, completing the spin interferometer on the return passage through the anticrossing. The probability of the \Dpit~state occupation, P$_{\Dpit}$, is directly obtained by this projection back into the (2,0,1) [or (1,0,2)] regime, where the required spin-to-charge information conversion is achieved by the Pauli Blockade \cite{Ono2002} of the \Qtt~state. An experiment with a double anticrossing (DA) pulse is also illustrated in \subfig{1}{c}. The sequence is similar with the important distinction that a larger pulse enables LZ tunneling processes through both anticrossings before again projecting back in the (2,0,1) regime having passed through both anticrossings twice. Important calibration information is obtained if the pulse time is longer than the coherence time, i.e.~$\tau$$>$\Tts, where the mixing at the \Dpit-\Qtt~anticrossing is detected independently of coherence effects.  Figure~\ref{fig:1}(d) plots this against magnetic field for a 9-mV-wide (1,1,1) regime midway between the narrow and wide (1,1,1) regimes. The two anticrossings form a ``spin arch'' which is used to extract the coupling parameters for the model. 

The distinction between our two regimes is now clear. In the case of a wide (1,1,1) region, close to zero detuning, both $J_{LC}$ and $J_{RC}\sim0$, so E$_{\Dpit}$$\approx$E$_{\Dit}$$\approx$E$_{\Qit}$. Away from zero detuning only two of the spins are coupled: right-centre (left-centre) at negative (positive) detuning. Experiments using DA pulses in this regime involve coupling to not only \Qtt~but also to \Qit. Thus this regime is not suitable for a two-level system involving three interacting spins. As a control experiment, however, in Fig.~\ref{fig:2} we plot the coherent LZS oscillations obtained in this regime for both positive and negative detuning with a SA pulse. These compare to the first LZS experimental results with DQDs from \cite{Petta2010} later described theoretically in \cite{Ribeiro2010, Sarkka2011}. The degree of LZ tunneling, i.e.~the relative size of $A$ and $B$ in the coherent  $A \left| \Dpit \right> + Be^{i\phi(t)} \left|\Qtt \right>$ state, depends upon the speed, v, through the anticrossing: P$_{LZ}$$=$$e^{-\frac{2\pi\Delta^2}{\hbar v}}$, where $2\Delta$ is the energy splitting at the anticrossing.  The visibility of the oscillations is a balance between this speed and \Tts. For an infinite \Tts, a rise time  $\sim$0.2~$\mu$s would produce a 50/50 superposition (see also \cite{Petta2010}). Experimentally it is found that a 6.6~ns pulse rise time (or 3.3~ns Gaussian time constant) leads to oscillations with the highest visibility. The \Tts, obtained from a single parameter fit to the data, ranges from 5 to 18~ns, consistent with previous DQD experiments where \Tts~was limited by fluctuations in the nuclear field environment~\cite{Petta2005}.

%Figure~\ref{fig:2}(b) plots both the experiment and model of the stability diagram in the presence of a fixed length LZS pulse. A sequence of stripes, i.e.~oscillations in P$_{\Dpit}$, as function of detuning  is observed parallel to the charge transfer line between (2,0,1) and (1,1,1). The stripes are also LZS oscillations which occur vs.~detuning for a fixed pulse length because of above E$_{\Dpit}$-E$_{\Qtt}$ dependence. 

%Here is Andy's new write-up for Col 2 of page 3:

In Figures~\ref{fig:3} and \ref{fig:4} we show results for experiments with DA pulses in a narrow (1,1,1) regime, where $J_{LC}$ and $J_{RC}$ are finite throughout and two well-defined qubit states exist between the two anticrossings (i.e.~simulations based on experimentally extracted parameters confirm that \Dpit~has moved far enough below the \Qit~state that no experimental features are related to interactions with the \Qit~state). The energy level diagrams for this regime are shown in \subfig{4}{a}. The stability diagram, measured in the presence of a fixed amplitude DA pulse at 25~mT, is shown in \subfig{3}{a}. The results reveal LZS resonances parallel to both charge transfer lines, consistent with theoretical simulations [\subfig{3}{b}] and confirming that coherence is maintained as the \Dpit~state is transformed from one dominated by coupling between left and centre spins to one dominated by right and centre spins, effectively demonstrating coherent pairwise exchange control. 

To gain further insight, \subfig{4}{b} and (c) show experimental and theoretical plots of the pulse duration dependence of LZS oscillations at different magnetic fields. Two boundaries marked with horizontal white dashed lines can be observed at fields above 25mT. The region between the boundaries corresponds to the regime between the two anticrossings, while the resonances correspond to LZS oscillations. It can be seen (e.g.~curved dotted lines) that the resonances double back on themselves. This is a direct observation of tracking the resonance across the maximum in the \Dpit~vs.~detuning curve [see \subfig{4}{a} and \subfig{1}{b}]. We speculate that operating at this spot may provide more protection from charge noise, as the energy levels become locally flat vs.~detuning. 

While the frequency of coherent oscillations grows with field, due to the increased spacing between the two qubit levels, it appears as if the experiment and theory differ by 20~mT for experimental data at 40~mT and by 15~mT for data at 25~mT. We attribute this to a dynamic nuclear polarization effect (DNP)~\cite{Brataas2011}. To make this quantitative we extract horizontal slices in \subfig{4}{b} at 40mT (blue and white lines) and fit them to obtain \Tts. The data are consistent with a 20~mT DNP effect. It is found experimentally that the values of \Tts~for the three-spin qubit experiments in \subfig{3}{c,d} (8 to 15~ns) are within error identical to the values from the two-spin qubit experiments. This is consistent with \Tts~being dominated by local uncorrelated nuclear field fluctuations since both sets of qubit states differ by the same total spin \cite{Baugh2006}. Finally we note that we also observe a resonance beyond the second anticrossing marked with a white spot in \subfig{4}{b}. This is a non-trivial feature corresponding to a resonance condition of two interacting spin interferometers, one between the two anticrossings and a second, beyond the second anticrossing.  

In conclusion, we have demonstrated coherent control of a qubit based on three-interacting-spin states. We have confirmed that there is no detectable change in the coherence time in the three-spin experiments compared to the two-spin experiments. We have realized the pairwise control of exchange for a three-spin system by pulsing the detuning energy of a triple quantum dot. The same technique should carry over when more quantum dots are added in series to increase the number of qubits. Pairwise control of exchange, as demonstrated here, will then be useful for building complex quantum algorithms based on electron spin qubits in quantum dots. 

\section{Acknowledgements}

\noindent We thank D.G. Austing, W. Coish, and E. Laird for discussions and O. Kodra for programming. A.S.S. and M.P.-L. acknowledge funding from NSERC. G.G., A.K, M.P.-L., and A.S.S. acknowledge funding from CIFAR. G.G. acknowledges funding from the NRC-CNRS collaboration.

\section{Author contributions}

\noindent Z.R.W. developed and grew the 2DEG heterostructure free of telegraphic noise;  A.K. fabricated the triple quantum dot device capable of reaching the few-electron regime;  P.Z., L.G., and S.A.S. designed and built the high frequency lines up to 50 GHz at milliKelvin temperatures;  P.Z., L.G., and G.G.  ran the cryogenic equipment; L.G., G.G., S.A.S., and M.P.-L. developed the pulsing techniques; G.G., L.G, and S.A.S. performed the measurements; G.G., L.G., and G.C.A. analysed the data; G.C.A. performed theoretical simulations; G.G. and A.S.S. wrote the manuscript and supplementary information with input from all authors; G.G., L.G., and G.C.A made the figures; G.G., L.G., G.C.A., S.A.S., M.P.-L., and A.S.S. participated in discussions concerning the experimental and theoretical results; and A.S.S. supervised the project.

%%%%%%%%%%%%%%%%%%%%%%%%%%%%
% We will insert here the old source for the supp. info.

%\newpage
\begin{center}
\underline{SUPPLEMENTARY INFORMATION}
\end{center}

\section{Methods and background}

The device is fabricated on a GaAs/AlGaAs heterostructure grown by molecular beam epitaxy with a density of 2.1$\times10^{11}$~cm$^{-2}$ and a mobility of 1.72$\times10^6$~cm$^2$/Vs. Ohmic contacts are used to contact the two-dimensional electron gas (2DEG) located 110~nm below the surface. TiAu gate electrodes are patterned by electron-beam lithography to allow electrostatic control of the triple quantum dot (TQD). Two gates are used to define quantum point contacts (QPCs) used as charge detectors on the left and right of the TQD. 

\begin{figure}[hbt]
\setlength{\unitlength}{1cm}
\begin{center}
\begin{picture}(5,5)(0,0)
\includegraphics[width=5cm, keepaspectratio=true]{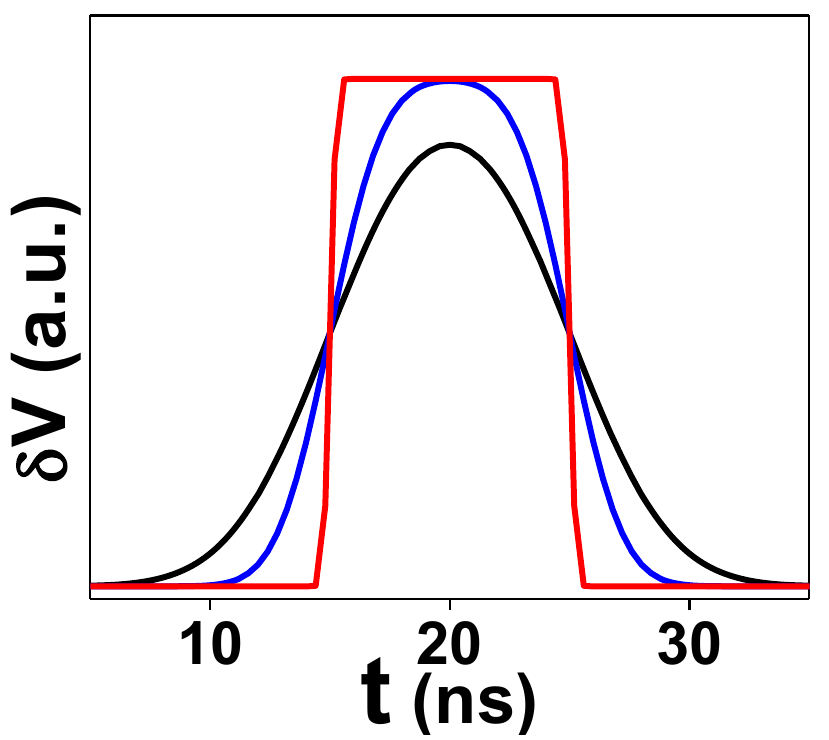}
\end{picture}
\end{center}
\caption{\footnotesize{Calculated pulse shapes for pulse duration $\tau$=10~ns after Gausian convolution, leading to rise times of 6.6, 3.5, and 0.4~ns.}}
\label{fig:SuppPulses}
\end{figure}

Charge detection measurements are made by measuring either the left or right QPC conductance with a lock-in technique using a typical root-mean-square modulation in the 0.05-0.1~mV range. The QPC detector conductance is tuned to below 0.1~e$^2$/h. High frequency pulses from two synchronized Tektronix AWG710B are applied via a bias-tee. The pulse of duration $\tau$ is typically $\leq$25~ns and the waveform is typically repeated every 2 to 5~$\mu$s. In most cases, the pulse rise times are controlled by passing the programmed rectangular pulses through low-pass filters internal to the AWG710B or through external Mini-Circuits SBLP filters. Typical rise times are 6.6~ns. In other cases, we use no filters, but we program pulses that are the convolution of a rectangular pulse with a Gaussian (see Fig.~\ref{fig:SuppPulses}). The details for the pulses used in the experiments are in Table~\ref{tab:Pulses}.

\begin{table*}[bht]
\begin{center}
\begin{tabular}{|c|c|c|c|c|c|c|c|}
\hline
Figure  & $\left|\epsilon_+-\epsilon_-\right|$ & ($\delta\Vone,\delta\Vtwo$) & Duration $\tau$  & Period T$_m$     & Rise time  & Filtered & Numerically convoluted     \\
		&		(mV)	&		(mV)				&	(ns)	&		($\mu$s)	&	 (ns)	&			& \\ \hline
1d & 9.0 &	(-8.8,11) & 16 & 2 & 6.6 & Yes & No \\ \hline
2a & 27  & (4.0,-1.7) & 1-16 & 2 & 6.6 & Yes & No \\ \hline
2c & 41.5 & (-4.11,7) & 1-16 & 2 & 6.6 & Yes & No \\ \hline
2b, 9a & ~50 & (4.0,-1.7) & 0-25 & 5 & 6.6 & Yes & No \\ \hline
2d, 9b & 27 & (-3.75,6.6) & 0-25 & 5 & 5.3 & Yes & No \\ \hline
3a & 5 & (-5.4,6) & 16 & 2 & 6.6 & No & Yes \\ \hline
3c,d, 4b (40~mT) & 5.6 & (-5,4.6) & 0-25 & 2 & 6.6 & No & Yes \\ \hline
4b (5~mT) & 3.9 & (-5.4.6) & 0-25 & 2 &  6.6 & No & Yes \\ \hline
4b (25~mT) & 5.1 & (-5.4.6) & 0-25 & 2 &  6.6 & No & Yes \\ \hline
4b (60~mT) & 4.6 & (-5.4.6) & 0-25 & 2 &  6.6 & No & Yes \\ \hline
7a, 8b (left) & ~50 & (4.0,-1.7) & 100 & 5 & 6.6 & Yes & No \\ \hline
7b, 8b (right) & 27 & (-3.75,6.6) & 100 & 5 & 3.3 & Yes & No \\ \hline
10 (top) & 24 & (4.0,-1.7) & 0-25 & 5 & 6.6 & Yes & No \\ \hline
11 & 34 & (-3.75,6.6) & - & 10 & 0.4 & No & No \\ \hline
12a,b & 9 & (-8,10) & 16 & 2 & 6.6 & Yes & No \\ \hline
12c & 9 & (-8,10) & 1-16 & 2 & 6.6 & Yes & No \\ \hline
13 & 9 & $\delta\Vone$=-0.8$\delta\Vtwo$ & 10 & 2 & 6.6 & Yes & No \\ \hline
\end{tabular}
\caption{\footnotesize{Pulse details for the experiments.}}
\label{tab:Pulses}
\end{center}
\end{table*}

The device is bias-cooled in a dilution refrigerator with 0.25~V on all gates. Once cold, suitable gate voltages are applied to the gates to form the TQD potential. 
%The dilution refrigerator has an electron temperature of $\sim$110~mK, as determined from the temperature dependence of the lineshape for an addition line.

\begin{figure*}[hbt]
\setlength{\unitlength}{1cm}
\begin{center}
\begin{picture}(16,4.5)(0,0)
\includegraphics[width=16cm, keepaspectratio=true]{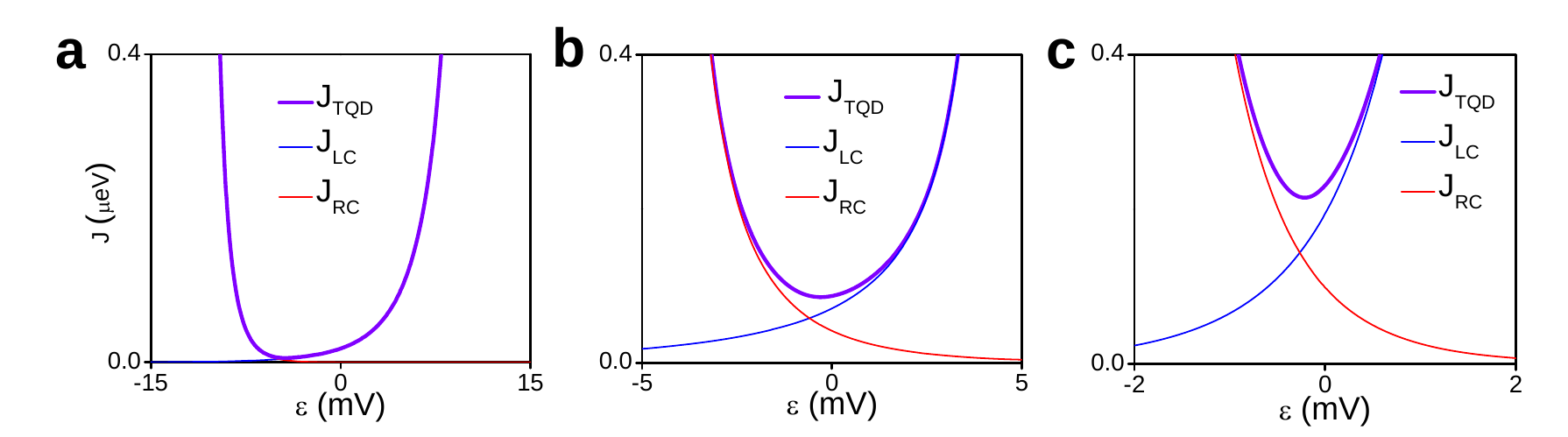}
\end{picture}
\end{center}
\caption{\footnotesize{Calculated energy diagrams showing the detuning dependence of $J_{TQD}$, $J_{LC}$, and $J_{RC}$. The exhange couplings increase when the size of the (1,1,1) region, $\left|\epsilon_+-\epsilon_-\right|$, decreases. (a) Wide (1,1,1) region: $\left|\epsilon_+-\epsilon_-\right|$=22~mV. (b) Medium (1,1,1) region: $\left|\epsilon_+-\epsilon_-\right|$=9~mV (c) Narrow (1,1,1) region: $\left|\epsilon_+-\epsilon_-\right|$=5.1~mV.}
\label{fig:JTQD}}
\end{figure*}

The charge detection stability diagram is shown in \subfig{MidSized}{a} for a 9-mV-wide (1,1,1) region. We focus on coherent spin manipulation between the three-electron spin states of the (1,0,2), (1,1,1), and (2,0,1) electronic charge configurations.  The dashed line illustrates a possible line of detuning, $\epsilon$. The detuning controls the energy difference between electronic configurations with the same total electron number that differ only by one charge transfer between adjacent dots. The detuning is increased by increasing \Vone~and decreasing \Vtwo~to stay along the dashed line in \subfig{MidSized}{a}. We express the detuning in millivolts along \Vone~and/or \Vtwo, but energies can be obtained by using the appropriate lever arms obtained from capacitance ratios and sweep angle in the stability diagram. At $\epsilon < \epsilon_-$, we have $\mu_{102} < \mu_{111} < \mu_{201}$, so (1,0,2) is the ground state [$\mu_{N_L N_C N_R}$ denotes the electrochemical potential of the (\NL,\NC,\NR) electronic configuration].  At $\epsilon = \epsilon_-$, we have $\mu_{102} = \mu_{111} < \mu_{201}$. This is the location of the stability diagram called the charge transfer line between (1,0,2) and (1,1,1), where a single charge is transferred from the right dot to the centre dot. When $\epsilon_- < \epsilon < \epsilon_+$, we have $\mu_{111} < \mu_{102}$ and $\mu_{111} < \mu_{201}$, so (1,1,1) is the ground state. At $\epsilon = \epsilon_+$, we have $\mu_{111} = \mu_{201} < \mu_{102}$, \textit{i.e.}~the charge transfer line between (1,1,1) and (2,0,1). Finally, at $\epsilon > \epsilon_+$, we have $\mu_{201} < \mu_{111} < \mu_{102}$, so (2,0,1) is the ground state. We assign $\epsilon$=0 to the middle of the (1,1,1) region, and we define the size of the (1,1,1) region as $|\epsilon_+-\epsilon_-|$ once projected onto the gate voltage axis on the same side as the QPC detector used in the measurement.

In the limit of large $\lvert \epsilon \rvert$, the ground state has a double electron occupation on one of the edge dots (left or right) and a spectator electron on the other edge dot (right or left). This double occupation reflects itself in a large singlet-triplet energy for the doubly occupied dot due to the onsite Coulomb repulsion and the orbital energy cost. If $\lvert \epsilon \rvert$ is reduced, the charge spreads via hybridization with the centre dot orbital due to the tunnel coupling between the edge dot and the centre dot. The Coulomb repulsion and the orbital energy cost both decrease, hence the smaller singlet-triplet energy difference when $\lvert \epsilon \rvert$ is small~\cite{Hanson2007}. We label the singlet-triplet energies for the two pairs of adjacent dots as $J_{LC}$ and $J_{RC}$. These are often called ``exchange energies.'' Their detuning dependence, assuming that significant charge hybridization is possible only between at most one pair of dots (which is valid for a large enough $\lvert \epsilon_+-\epsilon_- \rvert$), is given by Ref.~\cite{Laird2010} as
\begin{eqnarray} \frac{J_{LC}}{\tilde{\alpha}_{LC}}&=&(\epsilon-\epsilon_+)/2+\sqrt{[(\epsilon-\epsilon_+)/2]^2+\left(\frac{T_{LC}}{\tilde{\alpha}_{LC}}\right)^2} \label{JLC} \\
\frac{J_{RC}}{\tilde{\alpha}_{RC}}&=&(\epsilon_--\epsilon)/2 + \sqrt{[(\epsilon_--\epsilon)/2]^2+\left(\frac{T_{RC}}{\tilde{\alpha}_{RC}}\right)^2}
\label{JRC}
\end{eqnarray}
where $T_{LC}$ and $T_{RC}$ are the left-centre and right-centre interdot tunnel couplings respectively, and the $\tilde{\alpha}$'s are effective lever arms. Eqns.~(\ref{JLC}) and (\ref{JRC}) lead to $J_{LC}(\epsilon_+)=T_{LC}$ and $J_{RC}(\epsilon_-)=T_{RC}$, as expected. (We use the opposite convention for $\epsilon_+$ and $\epsilon_-$ as compared to Ref.~\cite{Laird2010}).

In cases where  $\lvert \epsilon_+-\epsilon_- \rvert$ is not large enough to approximate the system as two pairs of dots, we need to generalize the singlet-triplet energy by passing from the double dot, two-spin language where S, \Tplus, and \Tzero~play a role \cite{Petta2005, Petta2010} to the triple dot, three-spin language. Three spin-1/2 electrons can combine into quadruplets Q with total spin $S$=3/2 and doublets $\Delta'$ and $\Delta$ with $S$=1/2 \cite{Laird2010}. The generalized singlet-triplet energy that we are after is given by the energy difference between \Qit~and \Dpit. We label this energy as $J_{\mbox{TQD}}$ and its expression, reproduced from Ref.~\cite{Laird2010}, is given by
\begin{equation} J_{\mbox{TQD}}=\frac{J_{LC}+J_{RC}+\sqrt{J_{LC}^2+J_{RC}^2-J_{LC}J_{RC}}}{2}
\end{equation}
This expression has the expected two-spin limits if one of the exchange couplings is negligible. We show the calculated $J_{\mbox{TQD}}(\epsilon)$, along with $J_{LC}(\epsilon)$ and $J_{RC}(\epsilon)$ in Fig.~\ref{fig:JTQD} for the case of 22-, 9-, and 5.1-mV-wide (1,1,1) regimes.  

The LZS oscillations in the \Dpit~probability, P(\Dpit), are measured using standard spin-to-charge conversion techniques \cite{Ono2002}. Indeed, the QPC conductance measurement reveals whether the electronic configuration is (1,0,2), (2,0,1), or (1,1,1). If the measurement point is in the (2,0,1) region, the conductance G will be G$_{201}$ after long periods of time, as (2,0,1) is the ground state in this region. (We rely upon the finite relaxation time \Tone~in the spin-to-charge conversion regimes to achieve partial initialization.) Pulsing through the \Dpit-\Qtt~anticrossing in the (1,1,1) region for a duration $\tau$ allows the creation of a superposition of three-spin states. Immediately after the pulse, back in the (2,0,1) region, the system has a finite probability of being in a (1,1,1) charge state, such as \Qtt, \Qit, or \Dit. Without a spin flip, these states cannot make the charge transfer back to (2,0,1), as the energy cost would be too large. The only state that can get from (1,1,1) to (2,0,1) is \Dpit. The way the spin-to-charge conversion reveals P(\Dpit) is that \Dpit~corresponds to the (2,0,1) charge state at G=G$_{201}$, while \Qtt, \Qit, and \Dit~correspond to the (1,1,1) charge state at G=G$_{111}$. QPC conductance signals originating from an average over about a million pulses such that G$_{201}$$<$G$<$G$_{111}$ are linearly mapped to a finite P(\Dpit) between 1 and 0.

\section{Theoretical framework}

Based on Ref.~\cite{Laird2010}, the Hamiltonian for a system of three electron spins in the presence of a magnetic field along $\hat{z}$ is:

\begin{widetext}
\begin{equation}
H=J_{LC}\left(\vec{S_L}\cdot \vec{S_C}-\frac{1}{4}\right)+J_{RC}\left(\vec{S_R}\cdot \vec{S_C}-\frac{1}{4}\right)-E_z(S_L^z+S_C^z+S_R^z)
\end{equation}
\end{widetext}

where $J_{ij}$ is the exchange interaction between spins in dots $i$ and $j$, $\vec{S_i}$ is the spin in dot $i$, $E_Z$ is the Zeeman energy, and $\tilde{\alpha}_{LC}$ and $\tilde{\alpha}_{RC}$ are effective lever arms from capacitance ratios and pulse angle in the \Vone-\Vtwo~plane that allow the conversion from detuning $\epsilon$ in gate voltage units of mV into energy in $\mu$eV. According to Ref.~\cite{Laird2010}, the three-spin system is characterized by eight eigenvectors, which are divided into two subgroups by the exchange energy: four quadruplet states $Q$ with a total spin $S= 3/2$ ($S_z=\pm3/2, \pm1/2$) and two pairs of doublet states $\Delta$ and  $\Delta '$ with a total spin $S= 1/2$ ($S_z=\pm1/2$). 

%such as
%\begin{eqnarray}
%\frac{J_{LC}}{\tilde{\alpha}_{LC}}&=&(\epsilon-\epsilon_+)/2+\sqrt{[\epsilon-\epsilon_+)/2]^2+\left(\frac{T_{LC}}{\tilde{\alpha}_{LC}}\right)^2}  \label{eqn:JLC} \\ 
%\frac{J_{RC}}{\tilde{\alpha}_{RC}}&=&(\epsilon_--\epsilon)/2 + \sqrt{[\epsilon_--\epsilon)/2]^2+\left(\frac{T_{RC}}{\tilde{\alpha}_{RC}}\right)^2}
%\label{eqn:JRC}
%\end{eqnarray}

%In order to simplify the problem, we assume that the exchange energy $J_{LC}$ is negligible when the center and right dots are resonant and vice versa. This assumption is valid, if along the detuning line the charge transfer line between the left and center dots is very far from the charge transfer line between the center and right dots. In other words, when a pair of dots is resonant, the other pair has a large energy detuning, so the exchange for the latter pair is very small (In the double quantum dot approximation, the exchange energy between dots $i$ and $j$ along the detuning line is given by $J_{ij}(\varepsilon)=\frac{t_{ij}^2}{\varepsilon-\sqrt{\varepsilon^2+4E_{ij}^2}}$, where $t_{ij}$ is the tunnel coupling between the dots and $\varepsilon = 0$ when the dots are resonant~\cite{Taylor2007}). 
%We introduce the notation for the eigenstates identified by Ref.~\cite{Laird2010} for the case of our large positive detuning, but we drop the bar on the D states to simplify the notation, as swaping the left and right spins allows to readily get the states at large negative detuning:
We refer the reader to Ref.~\cite{Laird2010} for the complete list of eigenstates and eigenvalues, and we write down only those that play a role in the main text:

%\begin{eqnarray}
%\lvert Q_{-3/2}\rangle &=& \lvert\downarrow \downarrow \downarrow \rangle\\ 
%\lvert Q_{-1/2}\rangle &=& \frac{1}{\sqrt{3}}(\lvert\downarrow\downarrow
%\uparrow\rangle+\lvert\downarrow\uparrow\downarrow\rangle+\lvert\uparrow
%\downarrow\downarrow\rangle)\\
%\lvert Q_{1/2}\rangle &=& \frac{1}{\sqrt{3}}(\lvert\uparrow\uparrow
%\downarrow\rangle+\lvert\uparrow\downarrow\uparrow\rangle+\lvert
%\downarrow\uparrow\uparrow\rangle)\\
%\lvert Q_{3/2}\rangle &=& \lvert \uparrow \uparrow \uparrow \rangle \\ 
%\lvert {D}_{-1/2}\rangle &=& \frac{1}{\sqrt{6}}(\lvert\uparrow\downarrow\downarrow\rangle+\lvert\downarrow\uparrow\downarrow\rangle-2\lvert\downarrow\downarrow\uparrow\rangle)\\
%\lvert {D}_{1/2}\rangle &=& \frac{1}{\sqrt{6}}(\lvert\downarrow\uparrow\uparrow\rangle+\lvert\uparrow\downarrow\uparrow\rangle-2\lvert\uparrow\uparrow\downarrow\rangle)\\
%\lvert D'_{-1/2}\rangle &=& \frac{1}{\sqrt{2}}(\lvert\downarrow\uparrow\downarrow\rangle-\lvert\uparrow\downarrow\downarrow\rangle) \\
%\lvert D'_{1/2}\rangle &=& \frac{1}{\sqrt{2}}(\lvert\uparrow\downarrow\uparrow\rangle-\lvert\downarrow\uparrow\uparrow\rangle)\\
%\end{eqnarray}

\begin{widetext}
\begin{eqnarray}
\lvert Q_{1/2}\rangle &=& \frac{1}{\sqrt{3}}(\lvert\uparrow\uparrow
\downarrow\rangle+\lvert\uparrow\downarrow\uparrow\rangle+\lvert
\downarrow\uparrow\uparrow\rangle)\\
\lvert \Delta_{1/2}\rangle &=&
\frac{1}{\sqrt{4{\Omega}^2+2
\Omega(J_{LC}-2J_{RC})}}((J_{LC}-J_{RC}+\Omega)\lvert\uparrow\uparrow
\downarrow\rangle+(J_{RC}-\Omega)\lvert\uparrow\downarrow\uparrow
\rangle-J_{LC}\lvert\downarrow\uparrow\uparrow\rangle)\\
\lvert Q_{3/2}\rangle &=& \lvert \uparrow \uparrow \uparrow \rangle \\
\lvert \Delta'_{1/2}\rangle &=&
\frac{1}{\sqrt{4{\Omega}^2+2
\Omega(2J_{RC}-J_{LC})}}((-J_{LC}+J_{RC}+\Omega)\lvert\uparrow\uparrow
\downarrow\rangle-(J_{RC}+\Omega)\lvert\uparrow\downarrow\uparrow\rangle
+J_{LC}\lvert\downarrow\uparrow\uparrow\rangle)
\end{eqnarray}
%\end{widetext}

\noindent where $\Omega=\sqrt{J_{LC}^2+J_{RC}^2-J_{LC}J_{RC}}$ and the eigenvalues are:
%\end{widetext}
%\begin{widetext}
\begin{eqnarray}
E_{\Qit} &=& -E_Z/2\\
E_{\Dit} &=& -(J_{LC}+J_{RC}-\Omega+E_Z)/2\\
E_{\Qtt} &=& -3E_Z/2\\
%E_{\Delta} &=& -(J_{LC}+J_{RC}-\Omega)/2-E_ZS_{\Delta}^z\\
E_{\Dpit} &=& -(J_{LC}+J_{RC}+\Omega+E_Z)/2
\end{eqnarray}
\end{widetext}

\noindent In the limits of large $|\epsilon|$, the \Dpit~(\Dit) doublet state evolves to the $D'_{1/2}$ ($D_{1/2}$) and $\bar{D'}_{1/2}$ ($\bar{D}_{1/2}$) states of Ref.~\cite{Laird2010}, which involve a two-spin singlet or triplet plus a spectator (decoupled) spin-1/2.

%with eigenvalues:

%\begin{eqnarray}
%E_Q &=& -E_ZS_Q^z\\
%\end{eqnarray}

%\begin{eqnarray}
%\lvert \Delta_{-1/2}\rangle &=&
%\frac{1}{\sqrt{4{\Omega}^2+2
%\Omega(J_{LC}-2J_{RC})}}((J_{LC}-J_{RC}+\Omega)\lvert\downarrow
%\downarrow\uparrow\rangle+(J_{RC}-\Omega)\lvert\downarrow\uparrow
%\downarrow\rangle-J_{LC}\lvert\uparrow\downarrow\downarrow\rangle)\\
%\lvert \Delta_{1/2}\rangle &=&
%\frac{1}{\sqrt{4{\Omega}^2+2
%\Omega(J_{LC}-2J_{RC})}}((J_{LC}-J_{RC}+\Omega)\lvert\uparrow\uparrow
%\downarrow\rangle+(J_{RC}-\Omega)\lvert\uparrow\downarrow\uparrow
%\rangle-J_{LC}\lvert\downarrow\uparrow\uparrow\rangle)\\
%\lvert \Delta'_{-1/2}\rangle &=&
%\frac{1}{\sqrt{4{\Omega}^2+2
%\Omega(2J_{RC}-J_{LC})}}((-J_{LC}+J_{RC}+\Omega)\lvert\downarrow
%\downarrow\uparrow\rangle-(J_{RC}+\Omega)\lvert\downarrow\uparrow
%\downarrow\rangle+J_{LC}\lvert\uparrow\downarrow\downarrow\rangle)\\
%\lvert \Delta'_{1/2}\rangle &=&
%\frac{1}{\sqrt{4{\Omega}^2+2
%\Omega(2J_{RC}-J_{LC})}}((-J_{LC}+J_{RC}+\Omega)\lvert\uparrow\uparrow
%\downarrow\rangle-(J_{RC}+\Omega)\lvert\uparrow\downarrow\uparrow\rangle
%+J_{LC}\lvert\downarrow\uparrow\uparrow\rangle)\\
%\end{eqnarray}

Following the terminology of Ref.~\cite{Laird2010} the Hamiltonian for the Landau-Zener-St\"uckelberg (LZS) model in the \Dpit-\Qtt~system is:

\begin{equation}
\label{eqn:QDhamiltonian}
H=
\begin{pmatrix}
E_{\Qtt}& \Gamma_{{\Delta}',Q_{3/2}}\\
\Gamma_{{\Delta}',Q_{3/2}}^*& E_{\Dpit}
\end{pmatrix}
\end{equation}

\noindent where the off-diagonal term $\Gamma_{{\Delta}',Q_{3/2}}$ is the $\Dpit-\Qtt$ coupling originating from the hyperfine interaction betwen the electron spins and the nuclear spins via the $\hat{x}$ and $\hat{y}$ components of the Overhauser field gradients between the dots. This Hamiltonian is equivalent in the limit of weak ``spectator dot" coupling to the two-spin Hamiltonian in the $S$-$T_{+}$ basis described in Ref.~\cite{Taylor2007}.

For situations in which the ($\Qit$,$\Dit$) states play a role we use a Hamiltonian of the form: 

\begin{equation}
\label{eqn:interferencehamiltonian}
H=
\begin{pmatrix}
E_{\Qit}& \Gamma_{{\Delta},Q_{1/2}} & 0 & \Gamma_{{\Delta}',Q_{1/2}}\\
\Gamma_{{\Delta},Q_{1/2}}^* & E_{\Dit} &  0 & 0 \\ 
0 & 0 & E_{\Qtt} & \Gamma_{{\Delta}',Q_{3/2}}\\ \Gamma_{{\Delta}',Q_{1/2}}^*& 0 &
\Gamma_{{\Delta}',Q_{3/2}}^* & E_{\Dpit} \end{pmatrix} 
\end{equation}

\noindent The ($\Gamma_{{\Delta}',Q_{1/2}}$,$\Gamma_{{\Delta},Q_{1/2}}$) couplings are due to the $\hat{z}$ component of the Overhauser field gradients between the dots. The coupling $\Gamma_{{\Delta}',\Delta}$ is set to zero for spin conservation. This Hamiltonian is equivalent in the limit of weak ``spectator dot" coupling to the two-spin Hamiltonian in the $S$-$T_{+}$-$T_{0}$ basis. 

Note that the magnitudes of the off-diagonal coupling elements are empirically fitted to the observed magnitude of the LZS oscillations. This has no significant effect on the period of the LZS oscillations. The couplings are typically $\sim$ 0.1-0.2~$\mu$eV (see Table~\ref{tab:1} for the numerical values used in the calculations).

The time evolution of the density matrix $\rho$ is calculated from the initial state at large detuning where probability $P_{\Dpit}$=1, as described by the following equation:

\begin{equation}
\frac{d\rho}{dt}=i\left[\rho , H/\hbar \right] \end{equation}

\noindent The solution of the time evolution of $\rho$ involves a series of differential equations solved numerically by the Runge-Kutta method.
To simulate decoherence effects appropriate off-diagonal terms are included in the derivative of the density matrix leading to exponential decay of the resulting oscillations. 
The pulse shape is simulated by the convolution of a rectangular pulse of length $\tau$ with a Gaussian $\frac{1}{\sqrt{2\pi}s}e^{-t^2/2s^2}$ where $s$ is the Gaussian time constant, which is approximately equal to half of the measured rise time from 10\% to 90\% (see Fig.~\ref{fig:SuppPulses}). The applied magnetic field and the nuclear field gradient from the difference in Overhauser fields are kept constant.
At the end of the pulse the density matrix in the (\Qit,\Dit,\Qtt,\Dpit) basis is projected back onto \Dpit~to obtain P$_{\Dpit}$.

\section{Mapping the wide (1,1,1) regime onto a two-spin experiment}
%\section{Stability diagrams with long $\tau$ pulses}

\begin{figure}[hbt]
\setlength{\unitlength}{1cm}
\begin{center}
\begin{picture}(8,4)(0,0)
\includegraphics[width=8cm, keepaspectratio=true]{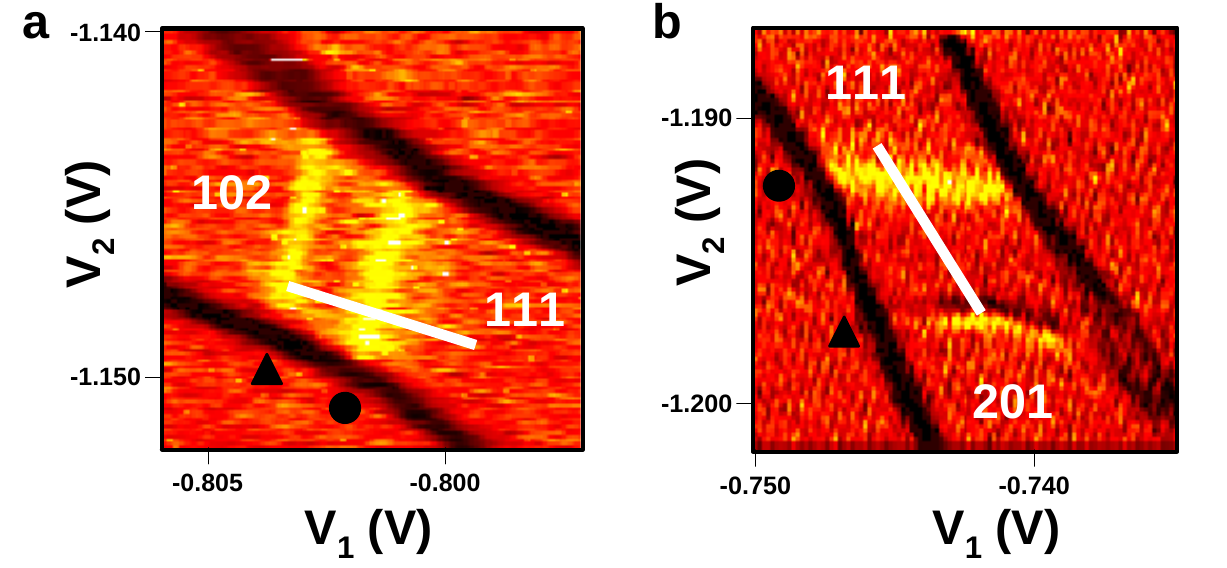}
\end{picture}
\end{center}
\caption{\footnotesize{(a) Numerical derivative of the right QPC conductance with respect to \Vone~in the presence of a pulse across the charge transfer line (black circle) between (1,0,2) and (1,1,1). The pulse is shown as a white line for a given (\Vone,\Vtwo) where signal is detected along the yellow line indicated by the black triangle when the end of the pulse reaches the \Dpit-\Qtt~anticrossing. Black is low, red is medium, and yellow is high transconductance. The pulse period is chosen $<\Tone$, which is $\sim$10~$\mu$s in this system (not shown). B=60~mT. (b) Numerical derivative of the left QPC conductance with respect to \Vtwo~in the presence of a pulse across the charge transfer line (black circle) between (2,0,1) and (1,1,1). The pulse is shown as a white line for a given (\Vone,\Vtwo) where signal is detected along the yellow line indicated by the black triangle when the end of the pulse is on the \Dpit-\Qtt~anticrossing. B=83~mT.}}
\label{fig:TwoSDs}
\end{figure}

For a pulse with long duration $\tau$=100~ns, a line, corresponding to the measurement location where the end of the pulse reaches the \Dpit-\Qtt~anticrossing, appears inside the stability diagram (black triangle), not far from the charge transfer line (black circle) inside the respective spin-blockade regions of (1,0,2) and (2,0,1) (Fig.~\ref{fig:TwoSDs}). We call this new line the \Dpit-\Qtt~line. In the absence of a pulse, we observe only the charge transfer line.

%\section{Magnetic field dependence and 2dot limit}

\begin{figure}[hbt]
\setlength{\unitlength}{1cm}
\begin{center}
\begin{picture}(8,11)(0,0)
\includegraphics[width=8cm, keepaspectratio=true]{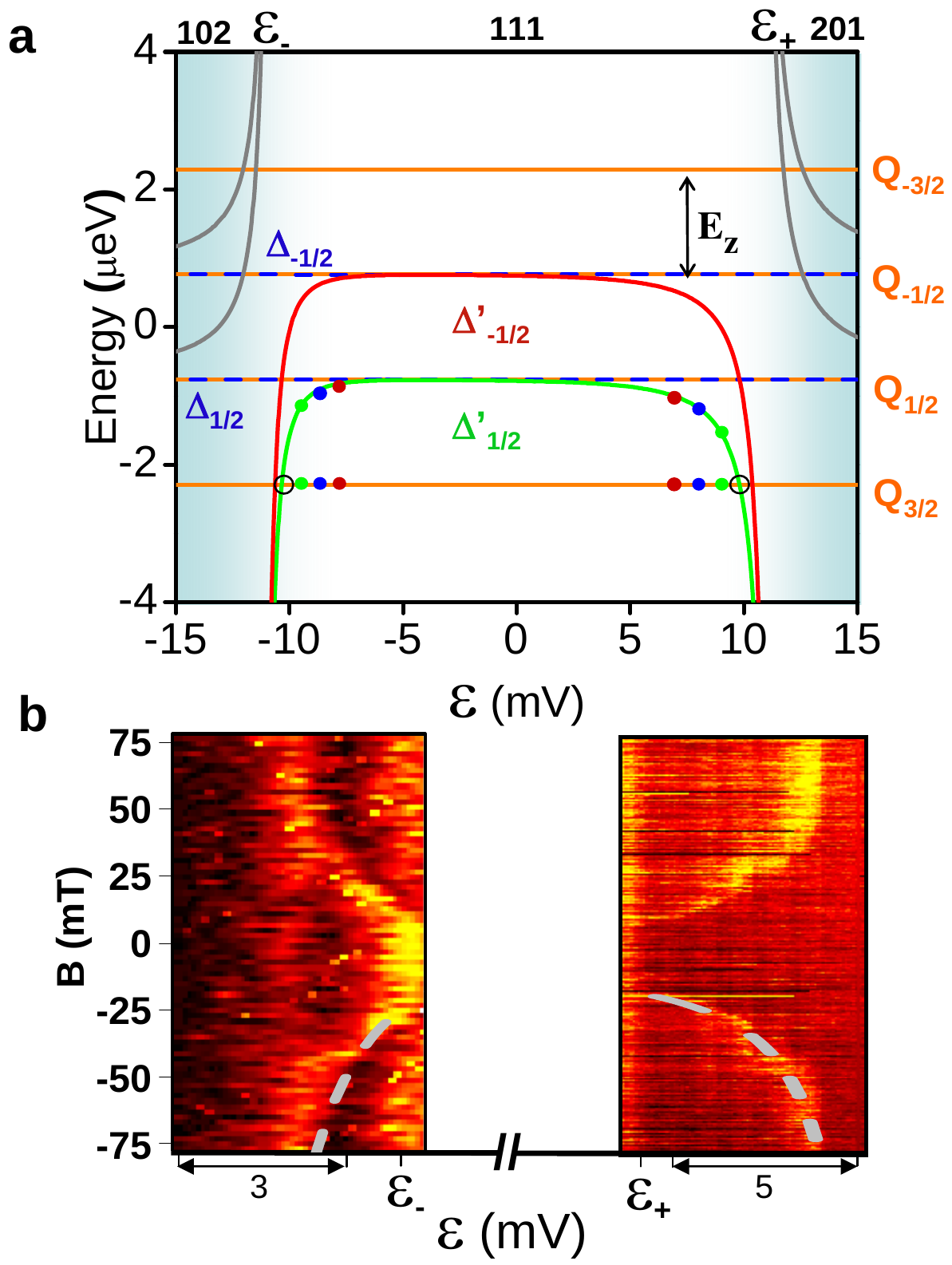}
\end{picture}
\end{center}
\caption{\footnotesize{Three-electron spin states energy diagram and spin funnels. (a) Calculated energies vs.~detuning $\epsilon$ for a 22-mV-wide (1,1,1) region, where three spin-1/2 electrons occupy the TQD in the presence of a Zeeman splitting E$_z$. The detuning line is at a \degrees{-45} angle with respect to the \Vone~axis in the \Vone-\Vtwo~plane. B=60~mT. The location of the \Dpit-\Qtt~anticrossings are indicated by black open circles. (b) Left [right] panel: spin funnel (half of the spin arch) in the numerical derivative of the right QPC conductance with respect to \Vone~[left QPC conductance (with a plane subtracted)] mapped in the detuning-B plane for the case of a wide (1,1,1) region. The pulse traverses the charge transfer line between (1,0,2) [(2,0,1)] and (1,1,1). The detuning axis is purely along \Vone~[\Vtwo]. The dashed lines are theoretical fits with detuning-dependent interdot couplings.}}
\label{fig:Supp2Funnels}
\end{figure}

The location of the two \Dpit-\Qtt~anticrossings depends on magnetic field, and the results are shown in \subfig{Supp2Funnels}{b}, where the individual spin funnels \cite{Petta2005} measured along two detuning axes combine to form a spin arch across the expanded detuning range. The energy diagram for the three-spin states in this case is in \subfig{Supp2Funnels}{a}.

To fit the spin arch we use two-parameter, detuning dependent couplings to 
generate $J_{LC}$ and $J_{RC}$ in Eqs.~(\ref{JLC}) and (\ref{JRC}). A simple constant coupling is 
found not to produce a good fit to the exchange couplings extracted from experiment, 
and this is corrected by an exponential multiplier such that

\begin{eqnarray}
T_{LC}(\epsilon) = 
\begin{cases}T_{LC}\mbox{exp}[C_{LC}( \epsilon - \epsilon_+)], & \epsilon <  \epsilon_+ \\
             T_{LC}, &  \epsilon \geq \epsilon_+
\end{cases}
\\
T_{RC}(\epsilon) = 
\begin{cases}T_{RC}\mbox{exp}[C_{RC}( \epsilon_- - \epsilon)], & \epsilon> \epsilon_- \\
              T_{RC}, &  \epsilon \leq \epsilon_- 
\end{cases}
\end{eqnarray}

\noindent The exponential form is used to ensure well-behaved functions away from the charge transfer lines. However, a linear fit works equally well in regions not too distant from the charge transfer lines.

\begin{figure}[hbt]
\setlength{\unitlength}{1cm}
\begin{center}
\begin{picture}(8,7.5)(0,0)
\includegraphics[width=8cm, keepaspectratio=true]{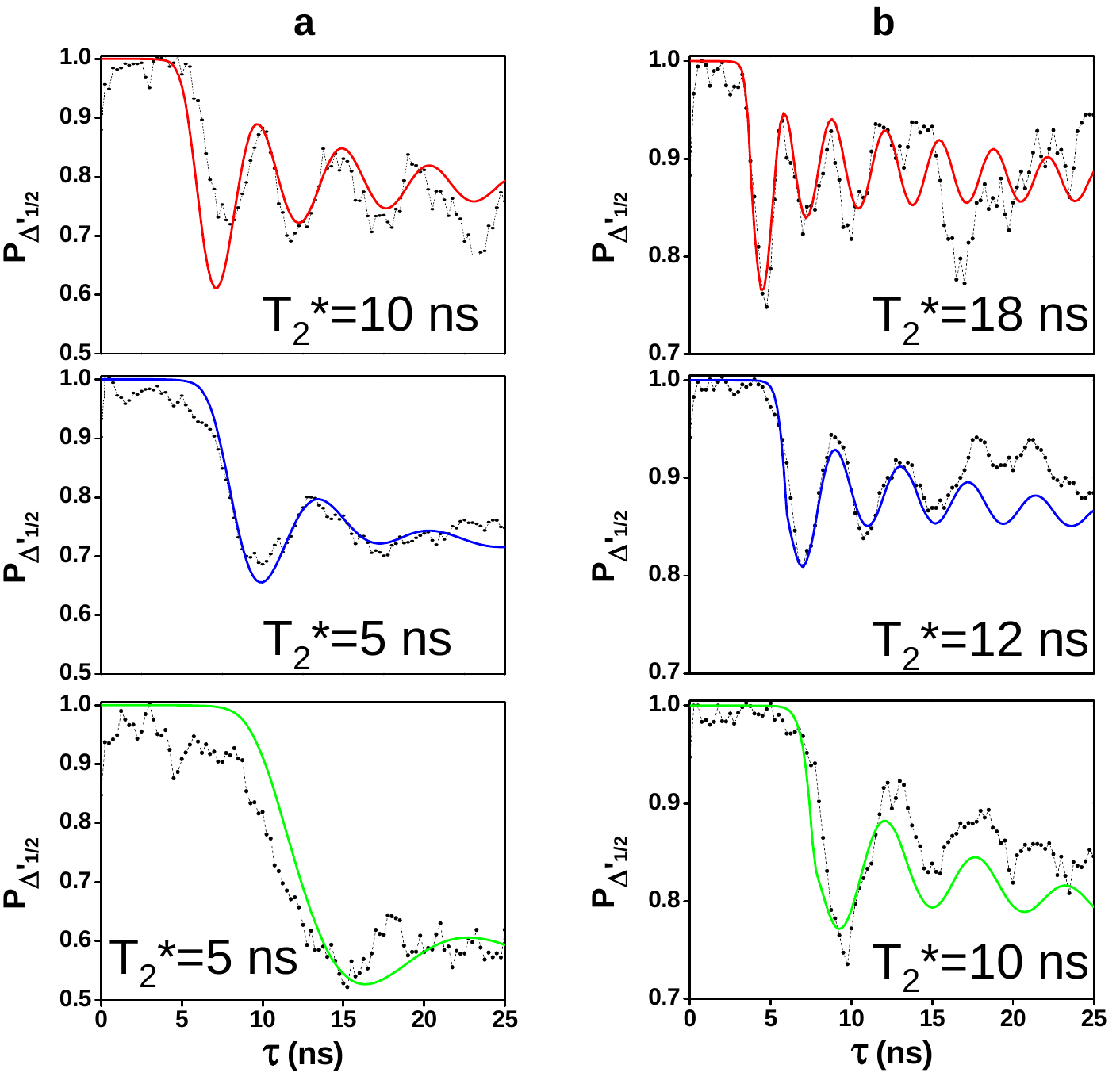}
\end{picture}
\end{center}
\caption{\footnotesize{Detuning dependence of the LZS oscillations of P$_{\Dpit}$ vs.~$\tau$ from experiments in the same conditions as in \subfig{Supp2Funnels}{b}. (a) [b] Pulses go from (1,0,2) [(2,0,1)] to (1,1,1). The experimental data are shown as points, while the lines for the theoretical fits for \Tts~at different detunings are colour-coded by the filled circles in \subfig{Supp2Funnels}{a}.} }
\label{fig:SuppDetuningDep}
\end{figure}

The detuning dependence of the LZS oscillations from the two \Dpit-\Qtt~qubits is shown in Fig.~\ref{fig:SuppDetuningDep}. The period of the LZS oscillations decreases as $|\epsilon|$ is increased, as expected. The single parameter fits for \Tts~reveal that it varies between 5 and 18~ns. 

\begin{figure*}[hbt]
\setlength{\unitlength}{1cm}
\begin{center}
\begin{picture}(16.5,5)(0,0)
\includegraphics[width=16.5cm, keepaspectratio=true]{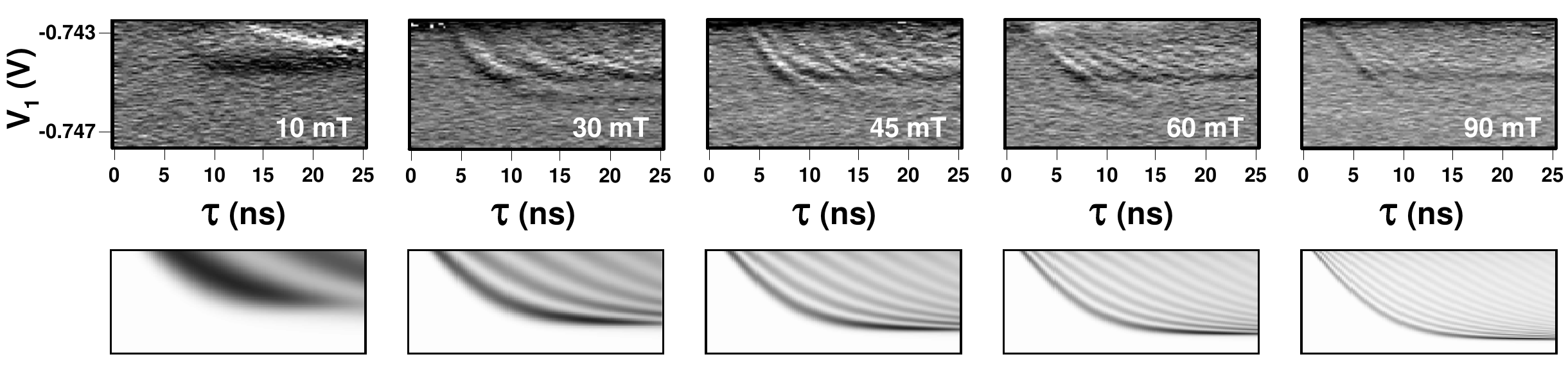}
\end{picture}
\end{center}
\caption{\footnotesize{Magnetic field dependence of the LZS oscillations. White (black) is low (high) transconductance. From left to right, B[mT]=10, 30, 45, 60, and 90. The top row contains the experimental maps in the $\tau$-detuning plane for the numerical derivative of the right QPC conductance with respect to detuning along \Vone. The pulse traverses the charge transfer line between (1,0,2) and (1,1,1). The detuning axis is purely along \Vone. The bottom row contains the corresponding maps of P$_{\Dpit}$ calculated from the LZS model of Eqn.~\ref{eqn:QDhamiltonian}. White (black) is high (low) P$_{\Dpit}$.}}
\label{fig:SuppMagField}
\end{figure*}

%\begin{figure}[hbt]
%\setlength{\unitlength}{1cm}
%\begin{center}
%\begin{picture}(8,9.5)(0,0)
%\includegraphics[width=8cm, keepaspectratio=true]{SuppRiseTime}
%\end{picture}
%\end{center}
%\caption{Rise time dependence of LZS oscillations. (a) Calculated pulse shapes for pulse duration $\tau$=10~ns after Gaussian filtering, leading to rise times of 6.6, 3.5, and 0.4~ns. (b-d) Left panel: experimental maps in the $\tau$-\Vone~plane from the numerical derivative of the right QPC conductance with respect to detuning along \Vone~for the pulse rise times corresponding to the colour code in (a). The pulse involves ($\delta V_1$,$\delta V_2$)=(4.0,-1.7)~mV to traverse the charge transfer line between (1,0,2) and (1,1,1) and repeats every $\tau_m$=5~$\mu$s. B=45~mT. Black is low, orange is medium, and yellow is high transconductance. Right panel: calculated P$_{\Dpit}$ maps in the $\tau$-\Vone~plane including the coupling to the (\Dit,\Qit)~states for the same rise times as in the left panel. Black (white) is low (high). Green [blue] dashed lines (guides to the eye) with negative [positive] slopes point to features pertaining to the \Dpit-\Qtt~[(\Dpit,\Dit)/\Qit] interaction.}
%\label{fig:SuppRiseTime}
%\end{figure}

\begin{table*}[bht]
\begin{center}
\begin{tabular}{|c|c|c|c|c|c|c|c|c|c|c|c|c|}
\hline
Figure  & $\left|\epsilon_+-\epsilon_-\right|$ & $J_{\mbox{TQD}}^{\mbox{min}}$ & $\frac{J_{LC}+J_{RC}}{2}$   &   $\tilde{\alpha}_{RC}$  &   T$_{RC}$   &       C$_{RC}$    &   $\tilde{\alpha}_{LC}$  &    T$_{LC}$   &      C$_{LC}$  &     $\Gamma_{\Delta ',\Qtt}$  &  $\Gamma_{\Delta ',\Qit}$ &    $\Gamma_{\Delta,\Qit}$ \\ 
        & (mV) &  ($\mu$eV) & ($\mu$eV) &   ($\frac{\mu\mbox{eV}}{\mbox{mV}}$)     &   ($\mu$eV)  &      ($\frac{1}{\mbox{mV}}$)   &     ($\frac{\mu\mbox{eV}}{\mbox{mV}}$)    &    ($\mu$eV)  &     ($\frac{1}{\mbox{mV}}$) &           ($\mu$eV)           &         ($\mu$eV)         &          ($\mu$eV)     \\ \hline  \hline

1d, 6b      &  9.0  & 0.116 & 0.0751 & 62.5       &    8.20   &     0.1627   &       38.0      &    5.28   &      0.061   &          -            &           -         &             - \\ \hline

1c, 12d,e,f, 13b      &  9.0  & 0.116 & 0.0751 & 62.5       &    8.20   &     0.1627   &       38.0      &    5.28   &      0.061   &          0.2            &           0.2         &             0.2 \\ \hline

4a (5 mT)     & 3.9 & 0.628   & 0.418 & 57.8     &      15.8 &       0.4995 &         39.0    &      13.4  &       0.380 &            -          &             -      &               - \\ \hline
4a (25 mT), 6c     & 5.1 & 0.309  & 0.191 & 57.8     &      15.8 &       0.4995 &         39.0    &      13.4  &       0.380 &            -          &             -       &               - \\ \hline
4a (40 mT)     & 5.6 & 0.229  & 0.140 & 57.8     &      15.8 &       0.4995 &         39.0    &      13.4  &       0.380 &            -          &             -       &               - \\ \hline

4a (60 mT)     & 4.6 & 0.394  & 0.262 & 57.8     &      15.8 &       0.4995 &         39.0    &      13.4  &       0.380 &            -         &             -       &               - \\ \hline

4c (5 mT)     & 3.9 & 0.628   & 0.418 & 57.8     &      15.8 &       0.4995 &         39.0    &      13.4  &       0.380 &            0.2          &             0.2       &               0.2 \\ \hline
3b, 4c (25 mT)     & 5.1 & 0.309  & 0.191 & 57.8     &      15.8 &       0.4995 &         39.0    &      13.4  &       0.380 &            0.2          &             0.2       &               0.2 \\ \hline
3c,d, 4c (40 mT)     & 5.6 & 0.229  & 0.140 & 57.8     &      15.8 &       0.4995 &         39.0    &      13.4  &       0.380 &            0.2          &             0.2       &               0.2 \\ \hline

4c (60 mT)     & 4.6 & 0.394  & 0.262 & 57.8     &      15.8 &       0.4995 &         39.0    &      13.4  &       0.380 &            0.2          &             0.2       &               0.2 \\ \hline

6a, 8a   &  22 & 0.0057 & 0.0037 &54.0       &    9.39  &       0.3414     &       40.0      &    9.96   &       0.1154  &             -             &            -          &             -   \\ \hline

8b(left)   &  $\sim$50 & - & - & 42.5       &    10.0  &      0.0   &       -      &    -   &      -    &         -           &         -         &            - \\ \hline
2b, 9a   &  $\sim$50 & - & - & 42.5       &    10.0  &      0.0   &       -      &    -   &      -    &          0.15           &          0.0          &            0.0  \\ \hline

8b(right)    &  27 & - & -  & -      &    -  &       -     &       35.9      &    5.89   &      0.0     &         -            &          -          &            - \\ \hline

2d, 9b(mid \& bottom)    &  27 & - & -  & -      &    -  &       -     &       35.9      &    9.96   &      0.1154     &         0.12            &          0.0          &            0.0 \\ \hline

9b(top)    &  27 & - & -  & -      &    -  &       -     &       35.9      &    9.96   &      0.1154     &         0.17            &          0.0          &            0.0 \\ \hline

10 (bottom)  & 24 & - & - & 42.5       &    9.39  &     0.3414    &      -       &  -     &    -  &          0.2              &         0.0           &          0.0   \\ \hline
\end{tabular}
\caption{\footnotesize{LZS model parameters. The column for ($J_{RC}$$+J_{LC}$)/2 is at the value of $\epsilon$ that gives the minimum of $J_{\mbox{TQD}}$.}}
\label{tab:1}
\end{center}
\end{table*}

%For $\tau$$<$25~ns we observe coherent LZS oscillations as a function of detuning and $\tau$, as shown in \subfig{LZS2Sides}{a,d}. The probability of measuring the \Dpit~state after the pulse, P$_{\Dpit}$, is plotted for two pulse detunings. From a single parameter fit, we extract values for \Tts, while the position of the fringes is completely determined by independently measured parameters indicated in Table~\ref{tab:1} following methods from Ref.~\cite{Petta2010}. The frequency of the LZS oscillations is directly related to the level spacing. This manifests itself in the data with a decreased frequency the closer the pulse is to the anticrossing, as expected. In general (not shown) such fits reveal that longer \Tts~values are obtained for pulses reaching further past the anticrossing. We believe such a behaviour would be consistent with charge noise effects due to the flattening of \Dpit~vs.~detuning.

Figure~\ref{fig:SuppMagField} contains the investigation of the magnetic field dependence of the LZS oscillations. The number of oscillations grows with B, as the energy difference between the two states in the qubit grows with E$_z$ due to the curvature in $\Delta '_{1/2}$. These experimental results compare very well to the calculations made with the single qubit model also shown in Fig.~\ref{fig:SuppMagField}. The truncation of the LZS oscillations along the \Vone~axis in the experimental data is due to a small spin-to-charge conversion region in this case, perhaps from a smaller singlet-triplet spacing for the right dot.

\begin{figure}[hbt]
\begin{center}
\includegraphics*[scale=0.7]{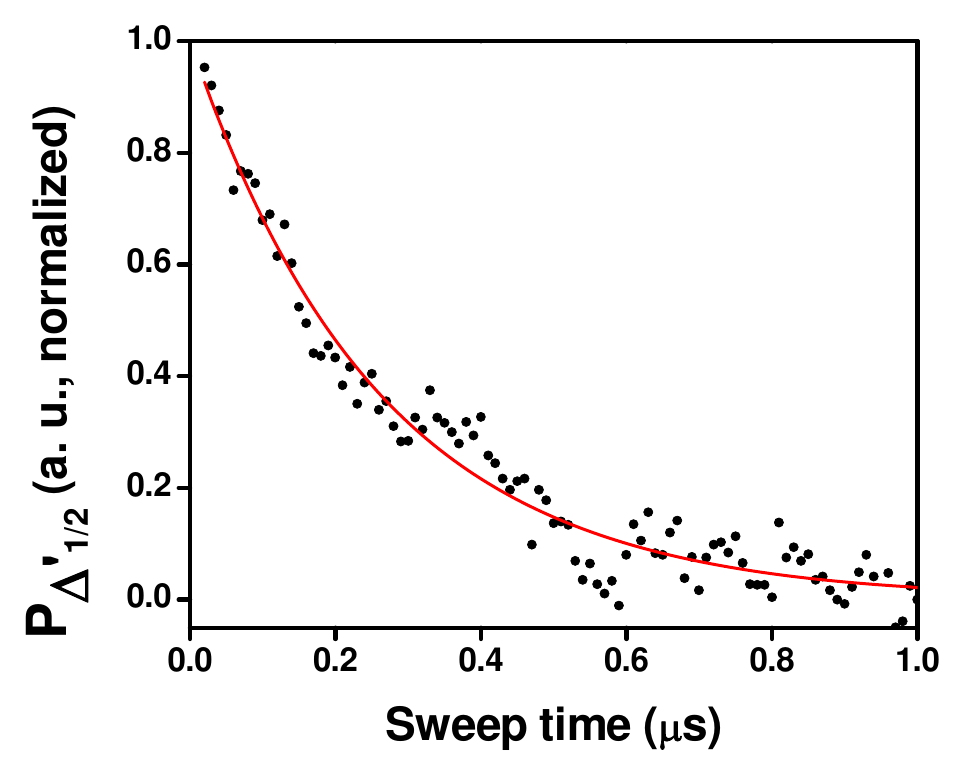}
\caption{\footnotesize{Landau-Zener transition probability for the avoided crossing between  \Dpit~and \Qtt~as a function of sweep time. The pulse traverses the charge transfer line between (2,0,1) and (1,1,1). The characteristic time extracted from the exponential fit is 260~ns.}}
\label{fig:PLZ}
\end{center}
\end{figure}

\begin{figure}[hbt]
\setlength{\unitlength}{1cm}
\begin{center}
\begin{picture}(8,10)(0,0)
\includegraphics[width=8cm, keepaspectratio=true]{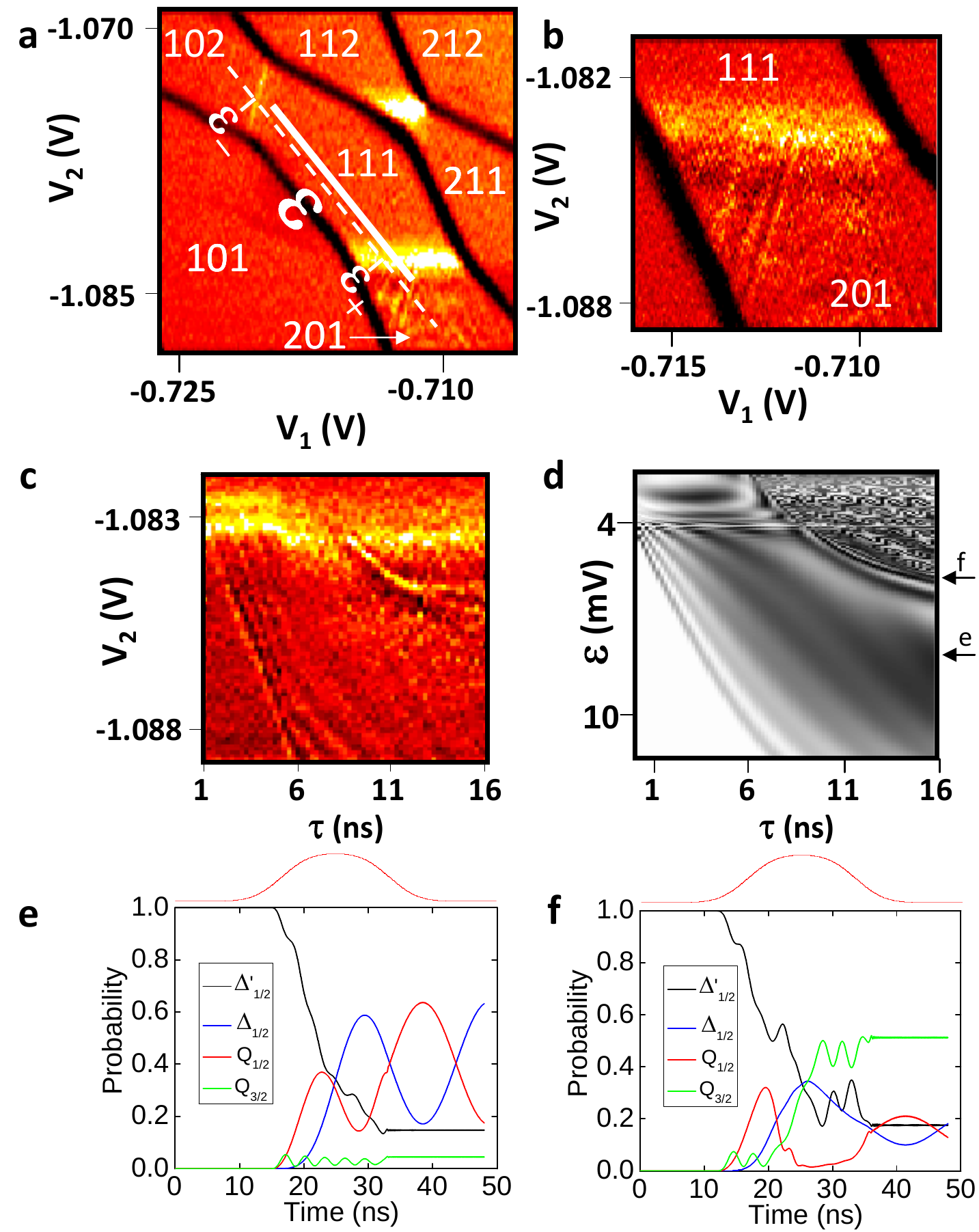}
\end{picture}
\end{center}
\caption{\footnotesize{Coherent three-spin state manipulation for a 9-mV-wide (1,1,1) region. (a) Stability diagram obtained from numerically differentiating the left QPC detector conductance with respect to \Vtwo~at B=60~mT. Black is low, orange is medium, and yellow is high. Charge addition lines appear black, and charge transfer lines appear yellow \cite{Granger2010}. A possible detuning line is drawn as a white dashed line. The pulse traverses the charge transfer line between (2,0,1) and (1,1,1) and reaches near the (1,0,2) charge transfer line. A pulse is drawn as a white line for a particular (\Vone,\Vtwo) where there is a signal when the end of the pulse reaches the far \Dpit-\Qtt~anticrossing. (b) Higher resolution experimental data zooming into the spin-to-charge conversion region (2,0,1) from (a) to show the details of the coherent features observed as lines parallel to the (1,0,2)-(1,1,1) charge tranfser line. (c) Experimental map in the $\tau$-\Vtwo~plane showing the coherent behaviour of the three-electron spin states from the numerical derivative of the left QPC conductance with respect to \Vtwo. \Vone~is swept proportionally to \Vtwo~in order to detune parallel to the pulse direction. The spin-to-charge conversion is performed in the (2,0,1) region for all the coherent oscillations observed in (c). (d) Calculated P$_{\Dpit}$ map in the $\tau$-$\epsilon$~plane for the same experimental settings as in (c). No dephasing is included to keep the fringes clearer. (e and f) Calculated probability of finding the system in each of the four indicated quantum states as a function of time before, during, and after the 16~ns pulse (shape shown above the graphs) for the two cases indicated by arrows in (d). Only the case shown in (f) has a reduced probability of ending in \Dit~and \Qit~and a large probability of ending in \Qtt.}}
\label{fig:MidSized}
\end{figure}

\begin{figure}[hbt]
\setlength{\unitlength}{1cm}
\begin{center}
\begin{picture}(8,4)(0,0)
\includegraphics[width=8cm, keepaspectratio=true]{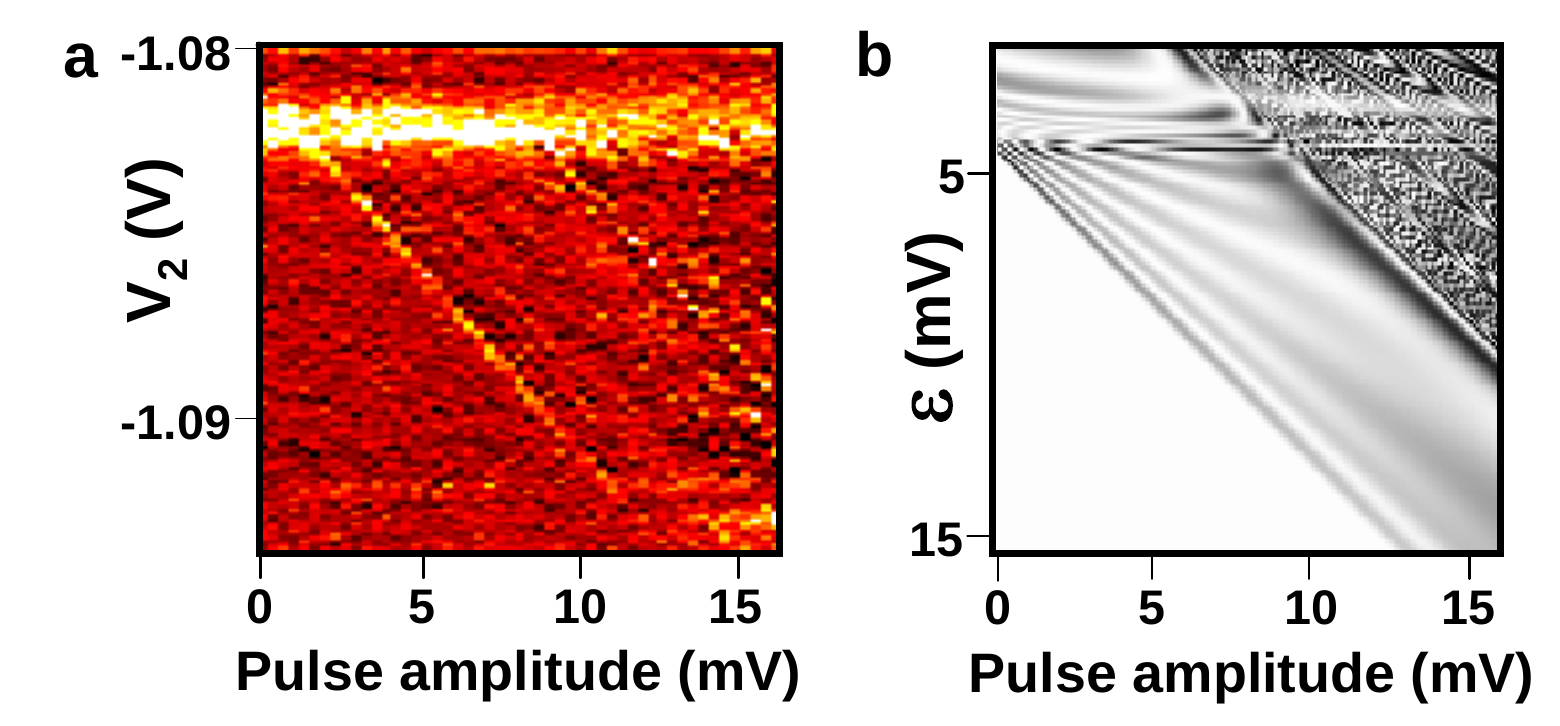}
\end{picture}
\end{center}
\caption{\footnotesize{(a) Experimental map in the (pulse amplitude)-\Vtwo~plane at $\tau$=10~ns revealing the coherent behaviour of the three-electron spin states from the numerical derivative of the left QPC conductance with respect to \Vtwo. The pulse traverses the charge transfer line between (2,0,1) and (1,1,1). \Vone~is swept proportionally to \Vtwo~in order to detune parallel to the pulse direction. The oblique yellow lines correspond to the two \Dpit-\Qtt~anticrossings. (b) Calculated P$_{\Dpit}$ maps in the (pulse amplitude)-$\epsilon$~plane for the same experimental settings as in (a).}}
\label{fig:AmplitudeDep}
\end{figure}

In order to measure the transition probability $P_{LZ}$ for the avoided crossing between states \Dpit~and \Qtt, we adopt the following procedure, originally used in a DQD system \cite{Petta2010}. From the (2,0,1) configuration, we induce the (1,1,1) transition by going non-adiabatically through the avoided crossing between the \Dpit~and \Qtt~in 0.5~ns. In this way, the  \Dpit~is preserved during the sweep. Immediately after that, we sweep through the avoided crossing in the opposite direction and measure the probability of returning in \Dpit, $P_{\Dpit}$, as a function of the return sweep time. Figure \ref{fig:PLZ} shows the results. When the sweep time for the return is short, \textit{i.e.}~in the non-adiabatic regime where $\Delta E/v\rightarrow 0$, Landau-Zener tunneling is efficient so $P_ {LZ}=1$ (see Ref.~\cite{Petta2010}). As the sweep time is increased, $P_{LZ}$ shrinks so the probability of finding the system in \Qtt~increases, which appears as an exponential decay of $P_{\Dpit}$. The characteristic time obtained by fitting an exponentially decaying function to the data is of 260~ns. This implies that to obtain a superposition of equal weights between \Dpit~and \Qtt~during a single sweep through the anticrossing, it is necessary to sweep through the transition for close to but less than 260~ns. This is impractical, as the decoherence time \Tts~is close to 15~ns \cite{Petta2010}.

\section{Mid-sized (1,1,1) region results}

For the 9-mV-wide (1,1,1) region of \subfig{MidSized}{a}, a set of oscillations parallel to the opposite charge transfer line (1,0,2)-(1,1,1) are observed in the (2,0,1) spin-to-charge conversion region when the pulse is large enough to engage the third spin in the coherent modulation by reaching the second (\textit{i.e.}~far) \Dpit-\Qtt~anticrossing. These new oscillations involve the coherent spin-state manipulation of three spins simultaneously. The zoomed-in version of these oscillations in the stability diagram is  shown in \subfig{MidSized}{b}. With a detuning line parallel to the pulse direction in the stability diagram (\textit{i.e.}~parallel to the centre dot addition line), it is possible to map out the coherent behaviour among the three-spin states, as seen in \subfig{MidSized}{c}. The set of closely spaced oscillations with negative slopes in the left part of the $\tau$-\Vtwo~plane corresponds to the LZS oscillations from the \Dpit-\Qtt~qubit that is close to the (2,0,1) region, while the narrow curved features in the right hand side of the $\tau$-\Vtwo~plane correspond to LZS oscillations involving the \Dpit-\Qtt~and (\Dpit,\Dit)-\Qit~interactions close to the (1,0,2) region. The corresponding theoretical P$_{\Dpit}$ map is shown in \subfig{MidSized}{d}. 

In order to get a better understanding of the fringes seen in the theoretical map of \subfig{MidSized}{d}, we plot in \subfig{MidSized}{e,f} the probabilities of finding the system in \Dpit, \Dit, \Qit, and \Qtt~as a function of time (before, during, and after the pulse) for the two fringes indicated by the arrows in \subfig{MidSized}{d}. Above these clear fringes is a dense group of fringes, where the relevant energy splittings grow sharply towards the (1,0,2) region producing very fast oscillations that are resolution limited by pixelation. The probability calculations at $\tau$=16~ns for the fringes corresponding to the arrows with the labels ``e'' and ``f'' in \subfig{MidSized}{d} are shown in \subfig{MidSized}{e} and (f), respectively. The broad fringe labelled ``e'' is mainly due to the (\Dpit,\Dit)-\Qit~interactions, as the probabilities of finding \Dit~and \Qit~are large. We stress that \Dpit~does not have a direct hyperfine coupling to \Dit~by spin conservation, so this is why the P(\Dit) lags compared to P(\Qit), as the weight in \Dit~depends on its interaction with \Qit. The well-defined fringe ``f'' from \subfig{MidSized}{d} is mainly due to the \Dpit-\Qtt~interaction. Indeed, the corresponding probability calculation in \subfig{MidSized}{f} reveals that P(\Qtt) is greater than P(\Qit) and P(\Dit), although these are not negligible. 

It is possible to decouple the $\tau$~dependence from the pulse amplitude changes that occur at small $\tau$ (reduction of the rectangular pulse amplitude by Gaussian convolution), by fixing the value of $\tau$ (e.g. 10~ns) and stepping the pulse amplitude. In such a map, the location of the two \Dpit-\Qtt~anticrossings are observed as two oblique yellow lines, seen in \subfig{AmplitudeDep}{a}. The corresponding calculated map of P$_{\Dpit}$ is shown in \subfig{AmplitudeDep}{b}. 

%\section{Narrow (1,1,1) region results}

%It is possible to minimize the interaction with \Dit~and \Qit~by shinking the (1,1,1) region to $\left|\epsilon_+-\epsilon_-\right|$$\sim$5~mV. The values of $J_{\mbox{TQD}}$ are larger than for the 9-mV-wide case, and we display that detuning dependence for the $\sim$5~mV wide region in \subfig{Narrow111}{a}. The calculated probability of finding any of the four states for $\tau$=16~ns corresponding to the red [blue] dot in \subfig{Narrow111}{b} is displayed in \subfig{Narrow111}{c} [\subfig{Narrow111}{d}]. A small mismatch between theory and experiment may be due to a nuclear pumping effect originating from the pulse, and such an effect is not taken into account in the calculation. From the calculated probabilities of finding the four quantum states, we find that P(\Dit) and P(\Qit) remain smaller than 10\%, so we deduce that the observed fringes for the case of $\left|\epsilon_+-\epsilon_-\right|$$\sim$5~mV originate from the two \Dpit-\Qtt~qubits.

%%%%%%%%%%%%%%%%%%%%%%%%%%%% 


\begin{thebibliography}{00}

%% \bibitem{label}
%% Text of bibliographic item

\bibitem{Petta2005} Petta, J. R. \textit{et al.} Coherent Manipulation of Coupled Electron Spins in Semiconductor Quantum Dots. \textit{Science} \textbf{309,} 2180--2184 (2005).

\bibitem{Koppens2006} Koppens, F. H. L. \textit{et al.} Driven coherent oscillations of a single electron spin in a quantum dot. \textit{Nature} \textbf{442,} 766--771 (2006).

\bibitem{Hanson2007} Hanson, R., Kouwenhoven, L. P., Petta, J. R., Tarucha, S., and Vandersypen, L. M. K. Spins in few-electron quantum dots.  \textit{Reviews of Modern Physics} \textbf{79,} 1217--1265 (2007).

\bibitem{PioroLadriere2008} Pioro-Ladri\`ere, M. \textit{et al.} Electrically driven single-electron spin resonance in a slanting Zeeman field.\textit{ Nature Phys.} \textbf{4,} 776--779 (2008).

\bibitem{DiVincenzo2000} DiVincenzo, D. P., Bacon, D., Kempe, J., Burkard, G, and Whaley, K. B. Universal quantum computation with the exchange interaction. \textit{Nature} \textbf{408,} 339--342 (2000).

\bibitem{Greentree2004} Greentree, A. D. \textit{et al.} Coherent electronic transfer in quantum dot systems using adiabatic passage. \textit{Phys. Rev. B} \textbf{70,} 235317 (2004).

\bibitem{Shevchenko2010}Shevchenko, S., Ashhab, S. and Nori, F. Landau-Zener-St\"uckelberg interferometry. \textit{Physics Reports} \textbf{492,} 1--30 (2010).

\bibitem{Zener1932} Zener, C. Non-Adiabatic Crossing of Energy Levels. \textit{Proc. R. Soc. Lond. A} \textbf{137,} 696--702 (1932).

\bibitem{Petta2010} Petta, J. R., Lu, H. and Gossard, A. C. A Coherent Beam Splitter for Electronic Spin States. \textit{Science} \textbf{327,} 669--672 (2010).

\bibitem{Loss1998} Loss, D. and DiVincenzo, D. P. Quantum computation with quantum dots. \textit{Phys. Rev. A} \textbf{57,} 120--126 (1998).

\bibitem{Ciorga2000} Ciorga, M.~\textit{et al.} Addition spectrum of a lateral dot from Coulomb and spin-blockade spectroscopy. \textit{Phys. Rev. B} \textbf{61,} R16315--R16318 (2000).

\bibitem{Elzermann2003} Elzermann, J. \textit{et al.} Few-electron quantum dot circuit with integrated charge read out. \textit{Phys. Rev. B} \textbf{67,} 161308(R) (2003).

\bibitem{Field1993} Field, M. \textit{et al.} Measurements of Coulomb blockade with a noninvasive voltage probe. \textit{Phys. Rev. Lett.} \textbf{70,} 1311--1314 (1993).

\bibitem{Laird2010} Laird, E. A. \textit{et al.} Coherent spin manipulation in an exchange-only qubit. \textit{Phys. Rev. B} \textbf{82,} 075403 (2010).

\bibitem{Granger2010} Granger, G. \textit{et al.} Three-dimensional transport diagram of a triple quantum dot. \textit{Phys. Rev. B} \textbf{82,} 075304 (2010). 

\bibitem{Taylor2007} Taylor, J. M. \textit{et al.} Relaxation, dephasing, and quantum control of electron spins in double quantum dots. \textit{Phys. Rev. B} \textbf{76,} 035315 (2007).

\bibitem{Ono2002} Ono, K., Austing, D. G., Tokura Y., and Tarucha S. Current Rectification by Pauli Exclusion in a Weakly Coupled Double Quantum Dot System. \textit{Science} \textbf{297,} 1313--1317 (2002).

\bibitem{Ribeiro2010} Ribeiro, H., Petta, J. R., and Burkard, G.  Harnessing the GaAs quantum dot nuclear spin bath for quantum control. \textit{Phys. Rev. B} \textbf{82,} 115445 (2010).

\bibitem{Sarkka2011} S\"arkk\"a, J. and Harju, A. Spin dynamics at the singlet�triplet crossings in a double quantum dot. \textit{New J. of Phys.} \textbf{13,} 043010 (2011).

\bibitem{Brataas2011} Brataas A. and Rashba E. Nuclear dynamics during Landau-Zener singlet-triplet transitions in double quantum dots. \textit{Phys. Rev. B} \textbf{84,} 045301 (2011).

\bibitem{Baugh2006} Baugh, J. \textit{et al.} Solid-state NMR three-qubit homonuclear system for quantum-information processing: Control and characterization. \textit{Phys. Rev. A} \textbf{73,} 022305 (2006). 

%\bibitem{Bluhm2010} H. Bluhm, S. Foletti, I. Neder, M. Rudner, D. Mahalu, V. Umansky, and A. Yacoby, arXiv:1005.2995v1 and references therein.

%\bibitem{Schroer2007} Schr\"oer, D. \textit{et al.} Electrostatically defined serial triple quantum dot charged with few electrons. \textit{Phys. Rev. B} \textbf{76,} 075306 (2007).

%\bibitem{Gaudreau2009} Gaudreau, L. \textit{et al.} A tunable few electron triple quantum dot. \textit{ Appl. Phys. Lett.} \textbf{95,} 193101 (2009).

%\bibitem{Rogge2009} Rogge, M. C. and Haug, R. J. The three dimensionality of triple quantum dot stability diagrams. \textit{New J. of Phys.} \textbf{11,} 113037 (2009).

%\bibitem{Amaha2009} Amaha S. \textit{et al.} Stability diagrams of laterally coupled triple vertical quantum dots in triangular arrangement. \textit{Appl. Phys. Lett.} \textbf{94,} 092103 (2009).

%\bibitem{Takakura2010} Takakura, T. \textit{et al.} Triple quantum dot device designed for three spin qubits. \textit{Appl. Phys. Lett.} \textbf{97,} 212104 (2010).

%\bibitem{Tanamoto2000} T. Tanamoto, Phys. Rev. A \textbf{61}, 022305 (2000).

%\bibitem{Saraga2003} D.S. Saraga and D. Loss, Phys. Rev. Lett. \textbf{90}, 166803 (2003).

%\bibitem{Fabian2005} J. Fabian and U. Hohenester, Phys. Rev. B \textbf{72}, 201304(R) (2005).

%\bibitem{Michaelis2006} B. Michaelis, C. Emary, and C.W.J. Beenakker, Europhys. Lett. \textbf{73}, 677 (2006).

%\bibitem{Busl2010} M. Busl, R. S\'anchez, and G. Platero, Phys. Rev. B \textbf{81}, 121306(R) (2010).

%\bibitem{Vernek2009} E. Vernek, C.A. B\"usser, G.B. Martins, E.V. Anda, N. Sandler, and S.E. Ulloa, Phys. Rev. B 80, 035119 (2009). 

%\bibitem{Rogge2009} M.C. Rogge and R.J. Haug, New J. of Phys. \textbf{11}, 113037 (2009).

%\bibitem{Gaudreau2006} L. Gaudreau, S.A. Studenikin, A.S. Sachrajda, P. Zawadzki, A. Kam, J. Lapointe, M. Korkusinski, and P. Hawrylak, Phys. Rev. Lett. \textbf{97}, 036807 (2006).

%\bibitem{Schroer2007} D. Schroer, A.D. Greentree, L. Gaudreau, K. Eberl, L.C.L. Hollenberg, J.P. Kotthaus, and S. Ludwig, Phys. Rev. B \textbf{76}, 075306 (2007).

%\bibitem{GaudreauPRB2009} L. Gaudreau, A.S. Sachrajda, S. Studenikin, A. Kam, F. Delgado, Y.P. Shim, M. Korkusinski, and P. Hawrylak, Phys. Rev. B \textbf{80}, 075415 (2009).

%\bibitem{Field1993} M. Field, C.G. Smith, M. Pepper, D.A. Ritchie, J.E.F. Frost, G.A.C Jones, and D.G. Hasko, Phys. Rev. Lett. \textbf{70}, 1311 (1993).

%\bibitem{Vidan2004} A. Vidan, R.M. Westervelt, M. Stopa, M. Hanson, and A.C. Gossard, Appl. Phys. Lett. \textbf{85}, 3602 (2004).

%\bibitem{footnote2} The exact values of C are slightly less negative here compared to the data in Figs.~\ref{fig:2} and \ref{fig:4} because of a background charge jump between the experiments.

%\bibitem{footnote3} Equivalently, one can think about transport of an electron at QP~6 from left to right through a cyclic permutation of the hole transport sequence starting with (1,1,2).

\end{thebibliography}
\end{document}